# HD 145263: Spectral Observations of Silica Debris Disk Formation via Extreme Space Weathering?

**C.M. Lisse[1], H.Y.A. Meng[2], M.L. Sitko[3], A. Morlok[4], B.C. Johnson[5], A.P. Jackson[6,7], R.J. Vervack Jr.[1], C.H. Chen[8], S.J. Wolk[9], M.D. Lucas[10,11], M. Marengo[10], D.T. Britt[12]**

[1]JHU-APL, 11100 Johns Hopkins Road, Laurel, MD 20723  carey.lisse@jhuapl.edu, ron.vervack@jhuapl.edu

[2]Steward Observatory, Department of Astronomy, University of Arizona, 933 N Cherry Ave, Tucson, AZ 85721 hyameng@gmail.com

[3]Department of Physics, University of Cincinnati, Cincinnati, OH 45221-0011 and Space Science Institute, Boulder, CO 80301, USA  sitkoml@ucmail.uc.edu

[4]Institut für Planetologie, Universität Münster, Wilhelm-Klemm-Straße 9 48149 Münster, Germany  morlokan@uni-muenster.de

[5]Department of Earth, Atmospheric, and Planetary Sciences, Purdue University, 550 Stadium Mall Drive, West Lafayette, IN 47904 bcjohnson@purdue.edu

[6]Centre for Planetary Sciences, University of Toronto, 1265 Military Trail, Toronto  ON, M1C 1A4, Canada

[7]School of Earth and Space Exploration, Arizona State University, 781 E Terrace Mall, Tempe, AZ 85287 ajackson@cita.utoronto.ca

[8]Space Telescope Science Institute, 3700 San Martin Dr. Baltimore, MD 21218  cchen@stsci.edu

[9]*Chandra* X-ray Center, Harvard-Smithsonian Center for Astrophysics, 60 Garden Street, Cambridge, MA, 02138 swolk@cfa.harvard.edu

[10]Department of Physics and Astronomy, 12 Physics Hall, Iowa State University, Ames, IA 50010 mmarengo@iastate.edu, mdlucas@iastate.edu

[11]Institute for Astronomy, University of Hawai'i at Manõa, 2680 Woodlawn Dr, Honolulu, HI 96822

[12]Department of Physics, University of Central Florida, Orlando, FL 32816 dbritt@**ucf**.edu

41 Pages, 5 Figures, 5 Tables

Proposed Running Title: **"HD 145263: Spectral Observations of Silica Debris Disk Formation via Extreme Space Weathering?"**

Please address all future correspondence, reviews, proofs, etc. to:

Dr. Carey M. Lisse

Planetary Exploration Group, Space Exploration Sector

Johns Hopkins University, Applied Physics Laboratory

SES/SRE, Bld 200, E206

11100 Johns Hopkins Rd

Laurel, MD 20723

240-228-0535 (office) / 240-228-8939 (fax)

Carey.Lisse@jhuapl.edu

# Abstract

We report here time domain infrared spectroscopy and optical photometry of the HD145263 silica-rich circumstellar disk system taken from 2003 through 2014. We find an F4V host star surrounded by a stable, massive $10^{22}$ - $10^{23}$ kg ($M_{Moon}$ to $M_{Mars}$) dust disk. No disk gas was detected, and the primary star was seen rotating with a rapid ~1.75 day period. After resolving a problem with previously reported observations, we find the silica, Mg-olivine, and Fe-pyroxene mineralogy of the dust disk to be stable throughout, and very unusual compared to the ferromagnesian silicates typically found in primordial and debris disks. By comparison with mid-infrared spectral features of primitive solar system dust, we explore the possibility that HD 145263's circumstellar dust mineralogy occurred with preferential destruction of Fe-bearing olivines, metal sulfides, and water ice in an initially comet-like mineral mix and their replacement by Fe-bearing pyroxenes, amorphous pyroxene, and silica. We reject models based on vaporizing optical stellar megaflares, aqueous alteration, or giant hypervelocity impacts as unable to produce the observed mineralogy. Scenarios involving unusually high Si abundances are at odds with the normal stellar absorption near-infrared feature strengths for Mg, Fe, and Si. Models involving intense space weathering of a thin surface patina via moderate (T < 1300 K) heating and energetic ion sputtering due to a stellar superflare from the F4V primary are consistent with the observations. The space weathered patina should be reddened, contain copious amounts of nanophase Fe, and should be transient on timescales of decades unless replenished.

# 1.    Introduction

Debris disks are the signposts of terrestrial planet formation and evolution (Zuckerman & Song 2004; Raymond *et al.* 2011). When giant planet accretion comes to a halt as the primordial disk gas dissipates after the first ~10 Myr of star formation (Mamajek 2009; Fedele *et al.* 2010; Ribas *et al.* 2015; Meng *et al.* 2017), solid materials in the remaining disk still grow by colliding and aggregating towards terrestrial planets (Kenyon & Bromley 2006; Chambers 2013; Morishima *et al.* 2010; Raymond *et al.* 2014). This process of "late stage" terrestrial planet accretion may last upwards of ~100 Myr, featuring large-scale, dust-producing impacts between planetesimals and planetary embryos (Kenyon & Bromley 2006, 2016; Genda *et al.* 2016). Fine dust produced in these large impacts can yield prominent spectral features observable in the mid-infrared. Silicate-emission debris disks have been observed around many young stars (Mittal *et al.* 2015); temporal variability has confirmed that recent large or giant impacts are indeed the culprits responsible for at least some of them (Meng *et al.* 2014, 2015).

The past decade has seen the emergence and development of the astromineralogy of dust with emission features based on *Spitzer*/IRS (Houck *et al.* 2004) observations (e.g., Lisse *et al.* 2006, 2007a, 2007b, 2008, 2009; Reach *et al.* 2009, 2010; Currie *et al.* 2011; Sitko *et al.* 2011; Olofsson *et al.* 2012; Fujiwara *et al.* 2010, 2012; Chen *et al.* 2014; Mittal *et al.* 2015) but also on data from large ground-based telescopes (e.g., Honda *et al.* 2004; Rhee *et al.* 2008). Mineralogical and chemical information is obtained by matching the linear superposition of laboratory model emissivity spectra to the *Spitzer* data. In the course of studying 20+ bright dust disk sources, we have noted a range of dust compositions, but find that the large majority of disks are composed of the Fe/Mg olivines, Ca/Fe/Mg pyroxenes, Fe/Mg sulfides, amorphous inorganic carbon, and water ice/gas, found in abundance in comets, asteroids, and KBOs in our solar system.

Lisse *et al.* 2009 examined the unusual silica-dust rich debris disk found in the HD 172555 system. Found surrounding a young (~23 Myr, Mamajek & Bell 2014) A7V star in the Beta Pic moving group, the mid-infrared spectrum of this system obtained by the Spitzer/IRS spectrometer in 2004 showed strong solid state emission features in the unusual 8–9 µm region, distinctly different from the usual 9 − 12 µm Si-O stretch features evinced by ferromagnesian olivines and pyroxenes. Analysis showed that these features were best matched by the emissivity spectra of amorphous silicas, like those found in terrestrial tektites, materials formed by flash re-freezing of rock vaporized in hypervelocity impacts. Given that the HD 172555 system contains no detectable circumstellar gas around the central A7V primary star, which is itself x-ray quiet, they argued that the only possible known mechanism for producing the observed silicas would be from a giant impact between two Mercury-sized (or larger) rocky planetesimals.

In this paper we discuss 5 mid-infrared spectral measurements of another system, HD 145263, an F4V star in the ~11 Myr old Upper Sco association (l = 349.7°, b = 18.73°, *Gaia DR2* distance = 142 pc; Preibisch & Mamajek 2006, Pecaut *et al.* 2012, Bailor-Jones *et al.* 2018) which also demonstrates strong 7–8 μm silica emission and 8–13 μm silicate emission features from warm (T ~ 285 K) dust at ~3 AU from the L = 4.1 $L_{solar}$ primary (Pecaut *et al.* 2012). HD 145263 is one of 7 well known debris disks in our NASA/IRTF northern sky Near InfraRed Disk Systems (NIRDS) survey with very large and structured excess flux at 8-13 μm due to thermal emission from warm (150–400 K) dust in the terrestrial planet zone of their systems (Fig. 1; total $F_{IR}/F_{bol}$ ~ 1x10$^{-3}$, Chen *et al.* 2011). Because of its low cluster age, it could harbor a proto-planetary or transition disk; as we find an optically thin disk with no evidence for circumstellar gas emission lines containing very processed dust, we can rule out a proto-planetary disk and must instead be observing a transition disk or very young debris disk. By simple scaling of its mid-infrared (hereafter MIR) flux, the ~0.4 Jy at 10 μm circumstellar disk must be massive, on the order of $10^{22}$ – $10^{23}$ kg (after allowing for their relative primary luminosities, it must be ~10x more massive than the 29 pc distant HD 172555, which is 1 Jy at 10 μm ).

Unlike HD 172555, however, we have 5 different MIR spectra taken in 2003 – 2014 to study (Fig. 1a), allowing us to search for temporal variability in the system. The first published mid-infrared HD 145263 spectrum, taken by Honda *et al.* 2004 using the ground base Subaru/COMICS instrument in 2003, did ***not*** show any evidence of silica dust emission features in our analysis, but instead, appeared to be a good match to the spectra found for dust emitted by active solar system comets. (At ~11 Myr age, it was plausible that the star has remnants of its primitive primordial birth disk still orbiting about it, especially since "The F-type members of Upper Sco appear to be all pre-main sequence" according to Pecaut *et al.* 2012.) The latter three spectra, taken using Spitzer/IRS in 2005, 2007, and 2009, appeared to show a markedly different composition than in 2003, one not reproducible using models of new dust replacement or high temperature thermal processing. At the same time the circumstellar disk emitted energy, particle size distribution (PSD), and temperature did not appear to have changed much, which initially led us to argue that selective *in situ* chemical processing of the circumstellar dust, driven by the central star, was at play.

As a check, we obtained later archival Subaru/COMICS spectra of HD 145263 from 2014, in order to secure that the apparent change from 2003 to 2005-2009 was real and not a relative calibration effect between the two different observatories. When we reduced the data, we found good agreement between the 2014 Subaru/COMICS 8–13 µm spectrum and the corresponding part of the Spitzer/IRS 2005–2009 spectra, obviating any calibration issues between the two datasets. However, when we applied the same data reduction methodology to the Subaru/COMICS 2003 data of Honda *et al.* as a further check, we could not reproduce the spectrum published in Honda *et al.* 2004, but instead found a spectrum consistent with the 4 other 2005 – 2014 spectra. We have thus obviated any rapid change occurring in the system's dust disk between 2003 and 2005. One of the most important goals for this paper is to set the literature record straight, in that what had appeared to be a pronounced change in the spectral character of the system's debris disk dust from 2003 to 2005 (in less than 2 years!) has become instead 11 years of a spectrally stable silica-rich debris disk.

Our initial work pursuing a rapid central-star driven metamorphosis of HD 145263's debris disk was not in vain, though, because in doing so we researched plausible non-impact driven physical mechanisms for producing a mixture of reduced, low temperature ferromagnesian pyroxenes + silica. One of these mechanisms, giant stellar XUV "space-weathering" alteration of dust particle surfaces in an initially primitive comet-dust like circumstellar population, matches very well with what is seen, and we thus report here mineralogical evidence for massively space-weathered circumstellar disk dust in the HD 145623 system. This was likely created by a giant primary star stellar flare (rather than an abnormally high stellar wind flux; at ~11 Myr age, the primary star is also likely still highly active and flaring), and if so, we predict that the effects of this space weathering should grind down on timescales of decades, revealing a more primitive, less processed mix of the "typical" ferromagnesian olivines + pyroxenes seen in young debris disks.

## 2. HD 145263 System Observations & Gross Characterization

The mid-infrared spectroscopic data we used for characterizing the HD 145263 system are the 2003 ground-based Subaru/Cooled Mid-Infrared Camera and Spectrometer (COMICS) 7.9–13.2 µm spectrum obtained by Honda *et al.* (2004), the archival 5.3–38 µm Spitzer/InfraRed Spectrograph (IRS) spectra obtained in 2005, 2007, and 2009, & the 2014 archival 7.9–13.2 µm Subaru/COMICS HD 145263 spectrum taken by Fujiwara *et al.* (2013). In the 2001-2009 timeframe, white light optical photometry from All-Sky Automated Survey (ASAS; Paczynski 2000; Pojmanski 2002; Pojmanski & Maciejewski 2004) and in white light by the Wide Angle Search for Planets (superWASP, Butters *et al.* 2010) was also obtained to monitor the system's variability. To bridge the two wavelength regimes, our group obtained $0.8 - 5.0$ µm spectra using the NASA/IRTF SpeX instrument in 2011-2013. For all the observations, the system was unresolved and the spectrophotometry represent the total flux emitted from the system of central star + circumstellar disk. The observational datasets used are summarized in Table 1.

### Table 1. Program Observations of the HD 145263 Debris Disk

| Source | Telescope/Instrument | AOR | λ (µm) | Observation Time (UTC) | BMJD_TDB | Mode |
|--------|---------------------|-----|--------|------------------------|----------|------|
| HD 145263 | ASAS | | 0.55 | 01/01/2001 – 10/01/2009 | | |
| | SuperWASP | | 0.55 | 07/05/2006 – 08/08/2008 | | |
| | Subaru/COMICS | | 7.9–13.2 | 15 Jul 2003 | 52835.3 | |
| | Subaru/COMICS | | 7.9–13.2 | 04 Jun 2014 | 56812.5 | |
| | *Spitzer*/IRS | 11196928 | 5.3–37.2 | 15 Mar 2005 | 53444.4 | SL+SH+LH |
| | *Spitzer*/IRS | 21808896 | 5.3–38 | 04 Sep 2007 | 54347.9 | SL+LL |
| | *Spitzer*/IRS | 26316288 | 5.3–38 | 18 Apr 2009 | 54939.6 | SL+LL |
| | ALLWISE | | 3.35, 4.60, 11, 22 | 24 Feb 2010 | 55251.5 | Photometric Skyscan |
| | IRTF/SpeX | | 2.4 – 5.0 | 21 Apr 2011 | 55672.5 | LXD |
| | | | 0.8 – 2.4 | 08 Jun 2013 | 56451.5 | SXD |
| | Spitzer/IRAC | | 3.55, 4.49 | 28 Apr 2013 -14 Nov 2013 | 56410.5 -56610.5 | Warm Spitzer Photometry |

**2.1 Spitzer Observations.** For the Spitzer/IRS 5.3–38 µm data, we first adopted the data products reduced by the optimal extraction in the CASSIS project (Lebouteiller *et al.* 2011, 2015), which utilize the latest Spitzer mission absolute calibration appropriate for point sources. The spectra then went through an iteration to reject the spectrophotometry at all wavelengths with signal-to-noise ratio below 3. To take care of the order-to-order calibration uncertainties, we forced the total fluxes over the overlapped wavelengths between adjacent orders to agree to the short wavelength values. Finally, the high-resolution spectrum in 2005 was smoothed to the wavelength grid of the low-resolution data. Since none of the spectroscopic observations has contemporary imaging by IRS Peak-up or MIPS, we did not additionally calibrate the spectra to any photometry. The homogeneous data reduction should best preserve the flux information and ensure the direct comparability of the results from different epochs.

**2.2 Subaru Observations.** To reduce the Subaru/COMICS 7.9–13.2 μm data, we used the specialized software package *q_series* (version 4.2) with identical procedures applied to both epoch's datasets. (Details of the reduction methodology can be found in the Subaru Data Reduction Cookbook: COMICS (Ray S. Furuya, 2012; https://subarutelescope.org/ Observing/DataReduction/Cookbooks/COMICS_ CookBook2p2E.pdf). HD145263 was observed twice by Subaru over 11 years, on July 15, 2003, and June 4, 2014, respectively. In addition to the spectroscopic data, photometric imaging observations were obtained at two wavebands with the F05C08.70W0.80 ($\lambda \sim 8.8$ μm) and F09C12.50W1.15 ($\lambda \sim 12.4$ μm) filters in 2003. In 2014, the same red filter (F09C12.50W1.15) was used with a different short-wavelength filter, F22C08.60W0.43 ($\lambda \sim 8.6$ μm).

For the spectra, we first averaged the dark field images and subtracted it from the dome flat field. The result was then convolved with a 20-pixel Gaussian kernel. The final flat field was made by dividing the convolved image by the original. Meanwhile, the averaged dark field was also removed from the object images. Next, the images were corrected for readout noise pattern and flat field, orthogonalized and fitted for a linear wavelength dispersion function, and used to subtract the residual sky pattern along the spatial axis. The outcomes were one science image for each pointing position with a positive spectrum from the odd chopping positions and a negative one from the even positions with an offset in the spatial axis. To measure the spectra, both the positive and the negative signals were extracted with an aperture width of 9 pixels. The noise level was estimated with two 9-pixel wide sky windows above and below the spectra, away from any image artifacts. The final spectrophotometric measurements were obtained as the weighted average of all extracted spectra of HD 145263.

The photometric images and standard star observations were prepared with regular dark subtraction and flat field correction. The chopping-nodding strategy commonly used in ground-based mid-IR observations should have, in theory, removed all sky photons. Therefore, we assumed sky brightness of 0 and adopted a radius of 6 pixels for aperture photometry. As a sanity check, we used a sky annulus of 15-30 pixels in radius around HD 145263 and found the average sky brightness fairly close to 0, with deviations slightly larger than the noise at a smaller scale (5 x 5 pixels), possibly due to some weak, medium-scale dark/flat field residual pattern. Following Honda *et al.* 2004, we fitted a linear equation for airmass and instrument efficiency effects by matching the observed spectra of the standard stars to their photometry with the two filters used in each observations. The resulting linear equation was then applied to the spectrum of HD 145263 for flux calibration. The calibrated Subaru spectra of HD 145263 in 2003 and 2014 are shown in Fig 1.

Our data reduction and calibration methodology produced a close match between the 2005-2009 Spitzer/IRS mission calibrated spectra and the 2003/2014 Subaru/COMICS spectra, giving us confidence in our analysis technique. What we didn't find was the marked difference between the 2003 Subaru/COMICS spectrum apparent from comparing the HD 145263 spectrum reported by Honda *et al.* 2004 and the archival Spitzer data, alleviating any need to explain a marked, rapid spectral change in the disk's spectral character in the 2 years between 2003 and 2005. In reconciling the qualitative discrepancy between the results from our reduction of the 2003 data, we have contacted Dr. Honda, who found an error in the stellar photometry used in the 2004 paper, which, if corrected, would lead to essentially the same result with our new re-reduction (M. Honda, private commun.)[1].

---

[1]To be fair, the stellar photometry for HD 145263 and nearby field stars has been updated since 2003, and the 2003 Subaru/COMICS data is of somewhat lower quality than the 2014 data. I.e., the total on-target integration time in 2003 is much shorter (2160 s vs 6664 s), leading to a slightly lower signal-to-noise ratio than in 2014. Also, no corresponding flat field was taken on the night HD 145263 was observed in 2003. We had to use the flat from the previous night, which provided less than ideal correction. Finally, 21 out of 24 spectroscopic images of HD 145263 in 2003 contained a bright ghost image. The position of the ghost moved from image to image, sweeping through a vast range in the spatial axis. To check the severity of the problem, we tried to only use the 3 ghost-free images but still got essentially the same spectrum. The ghost-contaminated images were valid in our case probably because we had visually inspected each image and carefully avoided the affected area with some margin when selecting the spectrum- and sky-extraction apertures. On the other hand, Watson *et al.* (2009), in performing their Spitzer/IRS study of PMS stars in Taurus, found that the shape of Spitzer 8-13 μm spectral features often differed from the Subaru/COMICS measurements, most likely due to problems observing with a warm telescope through the Earth's atmosphere,

**2.3 ASAS and superWASP Observations.** The five Subaru/COMICS + Spitzer/IRS spectra are shown together in Fig 1a, to demonstrate how stable the spectral character was between 2003 and 2014. At the same time, weekly long term photometric monitoring of HD 145263, in V band by the ASAS 10" telescopes in Las Camponas, Chile and Haleakala, HI  (Paczynski 2000; Pojmanski 2002; Pojmanski & Maciejewski 2004) and in white light by superWASP (Butters *et al.* 2010), detected no significant variation of HD 145263's brightness. The 2001–2009 V-band photometry for HD 145263 from the ASAS website (http://www.astrouw.edu.pl/ asas/?page=aasc) is reproduced in Fig 1b. There are no obvious significant brightening events seen that would be indicative of long-term stellar changes or intrasystem increases in stellar extinction. Whatever may have happened in the pre-2003 timeframe to alter the dust disk material, it must have occurred before 2001 or quickly enough between 2001 and 2003 to avoid ASAS detection, or had no detectable optical counterpart. Steady-state behavior of the system was also found in the 2013 Spitzer 3.6/4.5 μm photometric lightcurves of HD 145263 taken by Meng *et al.* 2015 (Table 1), who concluded that no detectable long-term temporal trends were occurring in the system during their 10-month long observational window.

**Table 2.    Properties of Star HD 145263 (IRAS F16078-2523; HIP 79288; SAO 184196)[a]**

| Name | Spectral Type | $T_\star$ (°K) | $M_\star$ ($M_\odot$) | $R_\star$ ($R_\odot$) | $L_\star$ ($L_\odot$) | $P_{rot}$ (days) | d (pc) | Age (Myr) | $f_{IR}/f_{bol}$[b] | $T_{dust}$ (K) | $r_{dust}$[c] (AU) | $v_{dust}$[d] (km/s) | $P_{orb,dust}$ (yrs) |
|---|---|---|---|---|---|---|---|---|---|---|---|---|---|
| HD 145263 | F4V (F3 to F5) | 6800 to 6500 | 1.45 to 1.36 | 1.58 to 1.49 | 4.5 to 3.7 | 1.74 to 1.75 | ~142 | ~11 | ~1x10⁻³ | 285 | ~3.0 | 20.6 | 4.1 |

[a] ZAMS stellar data from Pecaut *et al.* 2012, Mamajek 2018 (http://www.pas.rochester.edu/~emamajek/EEM_dwarf_UBVIJHK_colors_Teff.txt)
[b] From Chen *et al.* 2011, & estimated using $f_{IR}/f_{bol}$ = 1 x 10⁻³ for A7V HD17255, then scaling by 1 x 10⁻³ * (0.5Jy/1.0Jy)*(142pc/29pc)²
[c] Assuming $T_{dust}$ = 282 $L^{1/4}/r^{1/2}$
[d] For dust @ 3 AU and $M_\star$=1.4$M_\odot$,v=√(GM$_\star$/r)=√(G*1.4$M_\odot$/3AU)=√ (GM$_\odot$/2.1AU)=  29.8/√2.1 km/s = 20.6 km/s

Although the ASAS data had a longer time baseline on HD 145263 (689 measurements in 3192 days), the 2006 – 2008 superWASP monitoring was much more intensive (5512 measurements in 692 days) and more suitable for stellar periodicity analysis use. After filtering out nights with large internal measurement scatter inconsistent with the ASAS data, we used 1380 measurements and in estimating and removing the underlying continuum emission with the limited and noisy baseline available in the COMICS data. (The Watson group is also the source of the three CASSIS HD145263 Spitzer spectra used in this study.) Other groups have reported disagreements between IRS data and ground-based data (e.g. IRS vs BASS spectra, Kraus *et al.* 2013). However, our own more recent work with reducing Subaru/COMICS spectra for exodisks and comets has shown its calibration to be reasonable within ~15% relative spectral error, similar to the 10-15% error quoted for the Spitzer/IRS data.

in 25 good photometric nights over a baseline of 449 days and the SigSpec algorithm of Reegen (2007), a formal extension of the classical Lomb-Scargle, to analyze the filtered time series data. After ignoring observing cadence artifacts at 0.5, 1.0, and 6.67 days and aliases at 0.64 and 2.31 sdays, the periodogram (Figure 1c) shows a significant period of $1.747 \pm 0.001$ days with an amplitude of $0.005 \pm 0.001$ mag, reasonable values for a young, ~11 Myr old mid-F star.

If this quick rotational period reflects the equatorial rotation of HD 145263, then assuming a stellar radius of 1.55 $R_{Sun}$ suitable for an F4 dwarf (Table 2), we have a linear equatorial surface speed of v = 45 km/s. Based on their visual HD 145263 spectroscopy, Chen *et al.* (2011) measured a projected rotation speed for the star of v sin i = $40 \pm 3$ km/s. Equating these two estimates, we can solve for the sky inclination of the stellar equator, and find i = $63^{o\{+10\}}_{\{-8\}}$, (where i=$0^o$ is pole-on and i = $90^o$ is edge-on to the LOS). The HD 145263 circumstellar dust disk should have a similar inclination if it is coplanar with the primary's equator, implying that we are seeing the disk closer to edge-on than face-on, but still quite tilted upwards away from any problematic edge-on optically thick observing geometry. We should thus be able to observe changes in the disk's dust, if any, using unresolved spectroscopic measurements.

**2.4 SpeX Observations.** In 2011 and 2013 we returned to HD 145263 using the NASA IRTF/SpeX $0.8 - 5.2$ μm spectrometer, in order to further characterize the system as part of our 60+ object NIRDS exodisk/exoplanet spectral survey (Lisse *et al.* 2015, 2017a,b). The system was observed in two separate runs due to time limitations (with the longer $2.4 - 5.2$ μm integration performed first) but using the same stable A0V calibrator star for both runs. The IRTF/SpeX data are shown as a combined SED with the 2009 Spitzer/IRS data in Fig. 2a, and in detail in Figs 2b − 2e, to demonstrate that the primary star matches an F4V PHOENIX solar abundance stellar photospheric model (Husser *et al.* 2013) to high degree from 0.8–3.5 μm. At longer wavelengths ($3.5 - 5.0$ μm in the SpeX data, and from $5.3 - 37.2$ μm in the Spitzer/IRS data), the observed spectrum deviates above the PHOENIX continuum (i.e., demonstrates a 'flux excess') due to thermal emission from copious orbiting warm (T~ 285K) circumstellar dust. Based on our spectral classification experience with other NIRDS stars, a conservative range for the type of the HD 145263 primary is F3 − F5 V (i.e, F4 ±1 V). We use this conservative type range, along with the model stellar properties data published by Pecaut *et al.* 2012 and Mamajek 2018 (http://www.pas.rochester.edu/ ~emamajek/EEM_dwarf_UBVIJHK _colors_Teff.txt) to list the primary's properties in Table 2. It is worth noting that our SpeX F4V determination is in relatively good agreement with stellar properties for HD 145263 based on Gaia DR2 spectroscopic observations (Andrae *et al.* 2018): $T_{eff}$ = 6606 K, $A_G$ = 0.72 mag (extinction in Gaia G band, not V band), stellar radius 1.56 $R_{Sun}$, and luminosity 4.20 $L_{Sun}$.

Fig 2a also shows that of all the warm disk systems we have studied, the 8–13 μm "silicate emission complex" of HD 145263 best matches that of transition disk ~10 Myr old HD 166191 (Kennedy *et al.* 2014, Dore *et al.* 2016), strongly suggesting we are looking at transition disk behavioral effects in HD 145263. It is also interesting to note that the HD23514 and HD 15407A spectra form another group of similar silica-rich dust disk spectra, and HD 172555 yet another. We will return to the possibility that this implies multiple mechanisms for forming silica in circumstellar disks in Section 5.

Figs. 2b–e show that HD 145263 evinces no strong lines due to circumstellar CO gas at 2.3 – 2.5 μm, photospheric accretion (HBrγ or $H_2$ S(0)), outflowing stellar winds (HeI or FeII), or close-in sublimating dust. If HD 145263 is indeed a transition disk, it is a gas-poor or gas-free one.

## 3. Circumstellar Dust Mineralogical Modeling

## 3.1 Methodology

Spectral modeling was carried out by adopting a collection of possible mineralogical species and fitting various composite spectra to the observations. For an optically thin debris disk, a composite spectrum can be written as the combination of the contributions from individual species (Lisse *et al.* 2009),

$$F_\nu = \frac{1}{\Delta^2} \sum_i \int_{a_{min}}^{a_{max}} B_\nu [T_i(a, r)] Q_{abs,i}(a, \lambda) \pi a^2 \frac{dn_i(r)}{da} da$$

where $\Delta$ is the distance between the observer and the dust, $B_\nu$ is the Planck function at wavelength $\lambda$ and temperature $T$, which is dependent on the particle radius $a$ and the $i$-th composition at a distance $r$ from the central star, $Q_{abs}$ is the emission efficiency of the particle, and *dn/da* is the differential particle size distribution. Our spectral analysis consists of calculating the emission flux for a model collection of dust, and comparing the calculated flux to the observed flux. The emitted flux depends on the composition (location of spectral features), the particle size (feature to continuum contrast), and the particle temperature (relative strength of short versus long wavelength features). The contribution from scattered light is not considered as it is orders of magnitude weaker than dust emission in the mid-infrared (c.f. our measurements of the system's $0.8 - 35$ μm SED shown in Fig. 2a). The particle size distribution is usually assumed to be a power law or power law reduced at small particle sizes to model radiation pressure blowout effects.

The list of materials tested included multiple amorphous silicates with olivine-like and pyroxene-like composition; multiple ferromagnesian silicates (forsterite, fayalite, clino- and ortho-pyroxene, augite, anorthite, bronzite, diopside, and ferrosilite); silicas, both amorphous and crystalline (obsidian, tektites, quartz, cristobalite, tridymite); phyllosilicates (such as saponite, serpentine, smectite, montmorillonite, and chlorite); sulfates (such as gypsum, ferrosulfate, and magnesium sulfate); oxides (including various aluminas, spinels, hibonite, magnetite, and hematite); Mg/ Fe sulfides (including pyrrohtite, troilite, pyrite, and ningerite); carbonate minerals (including calcite, aragonite, dolomite, magnesite, and siderite); water-ice, clean and with carbon dioxide, carbon monoxide, methane, and ammonia clathrates; carbon dioxide ice; graphitic and amorphous carbon; and the neutral and ionized polycyclic aromatic hydrocarbon (PAH) emission models of Draine & Li (2007).

Our method has limited input assumptions, uses physically plausible laboratory emission measures from randomly oriented powders rather than theoretically derived values from models of highly idealized dust, and simultaneously minimizes the number of adjustable parameters. The free parameters of the model are the relative abundance of each detected mineral species, the temperature of the smallest particle of each mineral species, and the particle size distribution (PSD) power law slope (Table 3). Best fits are found by a direct search through (composition, temperature, PSD slope) phase space. Goodness-of-fit is determined using the reduced $\chi_v^2$ 95% confidence limit for the number of spectral degrees of freedom in each spectrum. (For more details of our spectral modeling analysis, we refer the reader to the descriptions given in Lisse *et al.* 2006, 2007a,b, 2008, 2009, and 2012; Reach *et al.* 2010; and Sitko *et al.* 2011).

**Table 3.    Composition of the Best-Fit Models[a] for the 2005, 2007, 2009 HD 145263 Spitzer Spectra**

| Species | Density $(g/cm^3)$ | Molecular Weight | Comet[b] Dust | HD145263 Weighted Surface Area[c] 2005 | 2007 | 2009 | Model[d] $T_{max}$ ($^o$K) | Model $\chi^2_\nu$ if species not included |
|---|---|---|---|---|---|---|---|---|
| **Detections** | | | | | | | | |
| ***Olivines*** | | | | | | | | |
| AmorphSil/Olivine Composition $(MgFeSiO_4)$ | 3.6 | 172 | 0.38 | 0.04 | 0.12 | 0.11 | 350 | 24.0/0.88/4.00/3.77 |
| Forsterite $(Mg_2SiO_4)$ | 3.2 | 140 | 0.23 | 0.26 | 0.23 | 0.22 | 350 | 1.91/1.88/2.49/2.49 |
| ***Pyroxenes*** | | | | | | | | |
| AmorphSil/Pyroxene Composition $(MgFeSi_2O_6)$ | 3.5 | 232 | 0.06 | 0.42 | 0.31 | 0.28 | 350 | 1.63/20.5/16.6/15.4 |
| FerroSilite $(Fe_2Si_2O_6)$ | 4.0 | 264 | 0.00 | 0.14 | 0.04 | 0.11 | 350 | 0.84/1.50/1.04/1.73 |
| ***Organics*** | | | | | | | | |
| Amorph Carbon (C) | 2.5 | 12 | 0.28 | 0.28 | 0.24 | 0.23 | 500 | 16.1/1.94/2.29/2.20 |
| ***Silicas*** | | | | | | | | |
| Tektite (Bediasite) (66% $SiO_2$, 14%$Al_2O_3$, 7% MgO, 6% CaO, 4% FeO, 2% $K_2O$, 1.5% $Na_2O$)[e] | 2.6 | 62 | 0.00 | 0.31 | 0.37 | 0.31 | 350 | 0.84/8.45/16.1/12.7 |
| SiO Gas (100% SiO) | N/A | 44 | 0.00 | 0.34 | 0.38 | 0.35 | 350 | 0.84/1.87/2.90/2.71 |
| **Upper Limits & Non-Detections[f]** | | | | | | | | |
| ***Silicas & Silicates*** | | | | | | | | |
| Amorphous Silica | 2.7 | 60 | 0.02 | 0.00 | 0.00 | 0.00 | 350 | 0.84/0.67/0.93/0.93 |
| Quartz | 2.65 | 60 | 0.00 | 0.00 | 0.00 | 0.00 | 350 | 0.84/0.67/0.93/0.93 |
| Ortho-Pyroxene $(Mg_2Si_2O_6)$ | 3.2 | 200 | 0.11 | 0.00 | 0.00 | 0.02 | 350 | 1.39/0.67/0.93/0.99 |
| Ortho-Pyroxene $(Mg_2Si_2O_6)$ | 3.2 | 200 | 0.11 | 0.00 | 0.00 | 0.02 | 350 | 1.39/0.67/0.93/0.99 |
| Diopside $(CaMgSi_2O_6)$ | 3.3 | 216 | 0.14 | 0.00 | 0.00 | 0.00 | 350 | 1.23/0.67/0.93/0.93 |
| ***Metal Sulfides*** | | | | | | | | |
| Ningerite (as $Mg_{0.10}Fe_{0.90}S$) | 4.5 | 84 | 0.10 | 0.00 | 0.00 | 0.00 | 350 | 3.78/0.00/0.00/0.00 |
| ***Water*** | | | | | | | | |
| Water Gas $(H_2O)$ | 1.0 | 18 | 0.08 | 0.00 | 0.00 | 0.00 | 210 | 0.87/0.67/0.93/0.93 |
| Water Ice $(H_2O)$ | 1.0 | 18 | 0.22 | 0.00 | 0.00 | 0.00 | 210 | 16.2/0.67/0.93/0.93 |
| ***Phyllosilicates*** | | | | | | | | |
| Smectite (Nontronite) $Na_{0.33}Fe_2(Si,Al)_4O_{10}(OH)_2 * 3H_2O$ | 2.3 | 496 | 0.00 | 0.00 | 0.01 | 0.00 | 350 | 0.84/0.67/0.97/0.93 |
| Talc $(Mg_3Si_4O_{10}(OH)_2)$ | 2.8 | 379 | 0.00 | 0.00 | 0.00 | 0.00 | 350 | 0.84/0.67/0.93/0.93 |
| Saponite | 2.3 | 480 | 0.01 | 0.00 | 0.00 | 0.00 | 350 | 0.87/0.67/0.93/0.93 |
| ***Carbonates*** | | | | | | | | |
| Magnesite $(MgCO_3)$ | 3.1 | 84 | 0.02 | 0.00 | 0.00 | 0.00 | 350 | 0.88/0.67/0.93/0.93 |
| Siderite $(FeCO_3)$ | 3.9 | 116 | 0.02 | 0.00 | 0.00 | 0.00 | 350 | 0.88/0.67/0.93/0.93 |
| ***Oxides*** | | | | | | | | |
| *Magnetite $(Fe_3O_4)$* | 5.2 | 232 | 0.00 | 0.00 | 0.00 | 0.04 | 350 | 0.84/0.67/0.93/1.45 |

(a) - Best-fit $\chi^2_\nu$ model with power law particle size distribution $dn/da \sim a^{-3.75}$, range of fit = 8-13 µm (2003), 6.3 – 34.7 µm (2005 – 2009). 95% C.L. has $\chi^2_\nu$ = 1.01/0.85/1.10/1.10.

(b) – Comet dust mineralogical composition as determined for the very primitive material of comet 73P/Schwassmann-Wachmann 3 as measured by Spitzer (Sitko *et al.* 2011).

(c) - Weight of the emissivity spectrum of each dust species required to match the HD 145263 emissivity spectrum, including a large rubble BB fit. Uncertainties on these are typically 0.02 (2-sigma), and weighted surface areas of any species at the ≤ 0.02 level are treated as non-detections.

(d) - All temperatures are ±10K (1σ). $T_{LTE}$ @ $r_h$ = 3.0 AU from the 4.1 $L_{solar}$ HD 145263 primary ~ 285 K. The best-fit temperature for the largest rubble grains, which act like a long wavelength cold blackbody function addition to the 5 – 15 µm portion of the spectrum, has $T_{bb}$ = 150 K, and amplitude = 2.1/0.9/0.9 in 2005/2007/2009.

(e) - Koeberl 1988

(f) – Upper Limits & Non-Detections : species which appear to improve the fit by eye, but do not improve the $\chi^2_\nu$ value above the 95% C.L. limit. These include crystalline silica, water, phyllosilicates, carbonates, and iron oxides, all expected products of aqueous alteration of rocky dust.

In Fig 3 we show how, using our spectral modeling, we can analyze the Spitzer observations and derive a dust mineralogy. (We focus on the Spitzer data as the longer wavelength baseline is much more constraining for the mineralogical makeup of HD 145263's circumstellar dust than the 8–13 μm Subaru/COMICS spectra, while the concordance of all 5 of the HD145623 mid-infrared spectra shown in Figure 1a shows that the Spitzer spectra are good proxies for the Subaru epochs as well.) We also present the Spitzer/IRS HD 172555 debris disk spectrum, in order to show how markedly qualitatively different it is in the sharpness of its emission peaks (due to ultrafine dust) and the lack of pyroxene features at 9.8 and 10.3 μm. *et al.* 2016). Figure 4 shows in graphical pie chart form the relative abundance of each dust species during the 3 Spitzer epochs of 5–35 μm spectroscopy.

## 3.2    Mineralogical Modeling Results

Model phase-space searching easily ruled out the presence of a vast majority of our library mineral species from the HD 145263 disk. Convincing evidence was found only for the following majority species in the 2005–2009 Spitzer spectra (Table 3, Fig 3): crystalline silicate forsterite, ferrosilite; amorphous silicates with pyroxene-like composition and a tiny amount of olivine-like composition; large amounts of tektite-like hydrated silica; and amorphous carbon. The fosterite, amorphous carbon, and dominant ($M = 10^{22}$–$10^{23}$ kg), cold (T~150K), optically thick rubble was found to be about the same at all epochs. After searching through phase space for the best-fit particle size distribution power law index, a single power law of dn/da ~ $a^{-3.75}$ was found to fit all the Spitzer epochs well. Thus a size distribution of fine dust similar to that found for the more active solar system comets was found to produce the strong mid-infrared spectral features seen, and the mass of this dust was roughly constant throughout.

The results of our mid-infrared spectral modeling of the three 2005 to 2009 HD 145263 Spitzer/IRS 5 – 35 μm spectra are shown in Figure 3 and 4 and are listed in Table 3 and 4. In each case we find dominant amorphous silica and amorphous pyroxene components, with about the same amount of amorphous carbon and crystalline Mg-olivine as in 2003. Very interestingly, all of the Fe-olivine fayalite, $FeMgS_x$, orthoenstatite, diopside, and water ice typically found in primitive cometary dust spectra (and by extension PPD disk dust; Bregman *et al.* 1987, Herter *et* *al.* 1987, Lisse *et al.* 2006, 2007a, Reach *et al.* 2010, Sitko *et al.* 2011) are missing from these spectra, as is 2/3 of the amorphous olivine, while large reservoirs of amorphous pyroxene and silica are present, along with some ferrosilite, to make a dust population quite unlike any we have ever studied before (including HD 172555, Fig 3d). The relative amount of fosterite, amorphous carbon, and silica found from 2005 to 2009 seems steady within the $\pm 0.02$ ($2\sigma$) noise of the spectral decomposition measurement, while there does seem to be a slight upward trend of amorphous silicate of olivine composition coupled with a slight downward trend of amorphous silicate of pyroxene composition.

A strong signature of 'SiO gas' is also present. However, since Lisse 2009, studies by Johnson *et al.* 2012 and Wilson *et al.* 2016 cast some doubt on the presence of this species, both from a photolytic stability and an actual measurement standpoint. The amorphous silicates/silica study of Speck *et al.* 2011, and the Mars Spirit rover study of Ruff *et al.* 2011 have demonstrated that this 8 μm shoulder feature may be a solid state feature found in hydrated forms of silica like hyalite (but not unhydrated forms like obsidian, cristobolite, or tektite). Our own 70+ IRTF/SpeX survey results of warm debris disks (Fig. 2) did not find any 4–5 μm SiO overtone emission in the 3 silica disk systems studied, including HD 145263. Importantly, the amplitude of the SiO gas feature is just as strong as the unhydrated tektite silica in each of our fits. We thus consider, at this point, that the "SiO gas feature" could be describing a hydrated silica solid state absorption feature, and could just as well be telling us about the amount of hydration of any silica present as it is about the presence of SiO gas in a system.

In Table 4 we present the derived atomic abundance counts in the detected disk dust vs that from a mix of comet-like dust. The H, C, O, Mg, Fe, S, and Ca abundances vs Si are markedly lower in the Spitzer spectra relative to that found in comet like dust, as expected for mechanisms involving evaporation of low temperature volatiles and sputtering of rocky materials by stellar wind energetic particles. Al and Na, which are more volatile rock-forming elements than Ca, are slightly higher in abundance in the 2005–2009 Spitzer spectra relative to comet-like dust.

**Table 4.  HD 145263 Circumstellar Dust vs. Cometary Dust Molar Atomic Abundances[a]**

| Species | Best-Fit Models Comet/2005/2007/2009 | Relative to Si = 1.0 Comet/2005/2007/2009 | Relative to Solar[b] Comet/2005/2007/2009 |
|---|---|---|---|
| H | 0.024/0.00/0.00/0.00 | 1.01/0.00/0.00/0.00 | 3.2e-5/0.00/0.00/0.00 |
| C | 0.058/0.05/0.05/0.048 | 2.40/1.64/1.64/1.57 | 0.32/0.22/0.22/0.21 |
| O | 0.100/0.096/0.096/0.10 | 4.11/3.1/3.1/3.3 | 0.17/0.13/0.13/0.14 |
| Si | 0.024/0.03/0.03/0.03 | 1.00/1.00/1.00/1.00 | 1.00/1.00/1.00/1.00 |
| Mg | 0.026/0.018/0.018/0.017 | 1.05/0.58/0.58/0.57 | 1.00/0.56/0.56/0.54 |
| Fe | 0.017/9.9e-3/9.9e-3/0.015 | 0.68/0.32/0.32/0.48 | 0.79/0.37/0.37/0.56 |
| S | 0.0053/0.00/0.00/0.00 | 0.22/0.00/0.00/0.00 | 0.51/0.00/0.00/0.00 |
| Ca | 0.0021/9.6e-5/9.6e-5/8.1e-5 | 0.088/3.2e-3/3.2e-3/2.6e-3 | 1.40/0.050/0.050/0.42 |
| Al | 0.000/4.3e-3/4.3e-3/3.6e-3 | 0.000/0.14/0.14/0.12 | 0.000/1.98/1.98/1.66 |
| Na | 0.000/0.00047/0.00047/0.00039 | 0.000/0.015/0.015/0.013 | 0.000/???/???/??? |

(a) – Molecular $N_{moles}(i) \sim$ Density(i)/Molecular Weight(i) * Weighted Surface Area (i) values given in Table 3, and atomic models are derived by summing up the atoms contained in each mineral species. Errors are ± 10% (1σ).  (b) Assuming the solar atomic abundances values vs Si are 3. 1 °ø 104 :7.6:14.1:1.00:1.05:0.87:0.43:0.063:0.072 (Anders & Grevesse, 1989; Asplund *et al.* 2005 ).

# 4.     Discussion

The goal of this paper is to use the information at hand to determine what could have formed the abundant silica found orbiting around HD 145263. Some quick conclusions can be made: In addition to the 5 epochs of mid-infrared spectral observations already presented, the 6[th] epoch includes an 0.8 − 5.0 μm IRTF/SpeX medium resolution spectrum to work with (Fig. 2), that immediately shows us that there is no detectable CO or $H_2O$ circumstellar gas in the system, so that flash heating by nebular shocks are unlikely; a much more sensitive search with ALMA for CO has also been negative (Lieman-Sifry *et al.* 2016). The IRTF/SpeX data also show that the Fe/Mg/Si stellar abundance ratios look normal and solar, so collisional grinding of silica-rich planetesimals are unlikely operant mechanisms for silica dust creation. (If the HD 145263 system was somehow extremely Si-rich compared to solar, we could possibly expect a felsic dust composition containing predominantly silica + feldspars.) Because the dust temperatures we derive from the SpeX + IRS SEDs (Figs 2, 5) is a moderate ~285K, far below the ferromagnesian silicate melting temperature of ~1000K, and because we find no detectable atomic sublimation emission features in our 2011 HD 145263 SpeX spectrum (like we saw for HR4796A, Lisse *et al.* 2017a), we can rule out close-in β Meteoroid formation via dust grain sublimation as a silica source.

Instead we focus on large-scale stochastic processes such as stellar flares (Osten & Wolk 2015) or giant impacts (Lisse *et al.* 2009) in HD 145263's past. We also need to consider that any debris disk in apparent steady state must be producing new dust to offset the removal of dust by radiation and stellar wind pressure and Poynting-Robertson drag, and that these sources have to be large, as the overall sensible surface area of the disk is on the order of 10 times that of the $10^{22}$ kg HD 172555 disk. So, unless the removal processes are very slow, whatever altered the orbiting dust population also seems to have altered its sourcing planetesimals.

This leads us to the following hypotheses for the production of silica in HD 145263's disk: **(1)** A giant impact involving large comet-like planetesimals destroyed all their highly volatile ices and metal sulfides, while reforming their less refractory rocky components into silica materials. **(2)** Visible light from a megascale stellar flare thermally flash-vaporized and re-processed all the orbiting dust and the surfaces of all sourcing grinding planetesimals. **(3)** A large stellar flare input enough energy to cause severe space weathering via low temperature heating + stellar wind sputtering of the surface layers of the circumstellar dust + sourcing grinding planetesimals.

Each of these hypotheses has pros and cons versus the known data for the system, as we lay out next. With the current data, we currently favor Hypothesis (3), and find fatal flaws with Hypotheses (1) & (2).

Before we go into detailed discussion of each of these model scenarios in turn, we compare in Table 5 what we know about the system's circumstellar dust properties versus other characterized disks. We see that it has some of the largest amounts ($M_{Moon}$ to $M_{Mars}$) of sensible warm, ~300K dust found in any non-primordial circumstellar disk. Given the system's young ~11 Myr old age and lack of circumstellar gas, the high disk mass is consistent with a transition disk system that is building planets or has just finished building planets (like HD11376 or HD166191; Lisse *et al.* 2008, Kennedy *et al.* 2014; Kenyon & Bromley 2006, 2016; Genda *et al.* 2016). We also note that compared to the circumstellar dust in other young, primitive, silicate-rich debris disks, **Mg**-olivine and amorphous carbon components are present at expected roughly comet/PPD levels, while amorphous and Fe-crystalline olivine have disappeared, apparently replaced instead with superabundant amorphous pyroxene and silica. The O:Si, C:Si, Mg:Si, and Fe:Si ratios are all lower than expected for primitive cometary material, showing evidence for selective O and Fe atom removal (Tables 3 & 4). This additional information will be important for discussing and evaluating the following formation hypotheses for the current state of HD145623's dust disk.

**Table 5. Derived Total Masses for Selected Relevant ExoDisks and Solar System Objects**

| Object | Observer Distance[1] (pc/AU) | Mean Temp[2] (K) | Equivalent Radius[3] (km) | 19 μm Flux[4] (Jy) | Approximate Mass[5] (kg) |
|---|---|---|---|---|---|
| Earth | --- | 282 | 6380 | | $6 \times 10^{24}$ |
| Mars | 1.5 AU | 228 | 3400 | | $6 \times 10^{23}$ |
| HD113766 (F2/F6) | 109 pc | 440 | $\geq 2700$ (270) | 1.85 | $\geq 2.3 \times 10^{23}$ ($5.3 \times 10^{20}$) |
| *HD145263/2005-9 (F4)* | *142 pc* | *285* | *$\geq 2180$ (1150)* | *0.49* | *>$1.8 \times 10^{23}$ ($3.5 \times 10^{21}$)* |
| ID8 (G8) | 357 pc | 900 | $\geq 2300$ (520) | 0.011 | $\geq 1.4 \times 10^{23}$ ($1.4 \times 10^{21}$) |
| Moon | 0.0026 AU | 282 | 1740 | | $7 \times 10^{22}$ |
| η Corvi (F2) Cold Dust | 18 pc | ~35 | $\geq 1600$ | | $\geq 6 \times 10^{22}$ |
| KBO Pluto | 40 AU | 45 | 1190 | | $1.3 \times 10^{22}$ |
| HD172555 (A5) | 29 pc | 335 | $\geq 910$ (250) | 0.90 | >$10^{22}$ ($1 \times 10^{20}$) |
| HD100546 (Be9V) | 103.4 pc | 250/135 | $\geq 910$ | 203 | $\geq 1 \times 10^{22}$ |
| Asteroid Belt | 0.1 - 5.0 AU | Variable | 577 | | $3 \times 10^{21}$ |
| Asteroid | 0.1 - 5.0 AU | Variable | 1 - 470 | | $1 \times 10^{13} - 9 \times 10^{20}$ |
| KBO Orcus | 30 - 45 AU | 42-52 | 300 | | $6 \times 10^{20}$ |
| Earth's Oceans | --- | 282 | 300 | | $6 \times 10^{20}$ |
| Enceladus | 10.5 AU | 75-135K | 252 | | $1 \times 10^{20}$ |
| Miranda | 19.2 AU | ~65 | 235 | | $7 \times 10^{19}$ |
| η Corvi (F2)Warm Dust | 18.2 pc | 400 | $\geq 900$ (140) | | $\geq 9 \times 10^{21}$  ($9 \times 10^{18}$) |
| KBO 1996 TO66 | 38 - 48 AU | 41 – 46 | ~200 | | ~$4 \times 10^{19}$ |
| Centaur Chiron | 9 – 19 AU | 65 – 95 | ~120 | | ~$1 \times 10^{19}$ |
| Outer Zody Cloud | 5 – 100 AU | 25 – 100 | 84 | | $4.9 \times 10^{18}$ |
| HD69830 (K0V) | 12.6  pc | 340 | $\geq 60$ (30) | 0.11 | $\geq 2 \times 10^{18}$ ($3 \times 10^{17}$) |
| Inner Zody Cloud | 0.1 - 5.0 AU | 260 | 9 - 14 | | $1 - 4 \times 10^{16}$ |
| Comet nucleus | 0.1 – 10 AU | Variable | 0.1 – 30 | | $10^{12} - 10^{15}$ |
| Hale-Bopp coma | 3.0 AU | 200 | | 144 | $2 \times 10^{9}$ |
| Tempel 1 ejecta | 1.51 AU | 340 | | 3.8 | $1 \times 10^{6}$ |

Red – Warm/hot dust; Aqua – liquid water; Blue – cold icy dust.
(1) - Distance from Observer to Object, updated for Gaia DR2 measurements (Bailor-Jones *et al.* 2018).
(2) - Mean temperature of thermally emitting surface.
(3) - Equivalent radius of solid body of 2.5 g $cm^{-3}$.
(4) - System or disk averaged flux.
(5) - Lower limits are conservative, assuming dust particles of size of 0.1 — 100 μm (as detected by *Spitzer*, in parentheses), or 0.1μm – 100m (extrapolated using the best-fit power law PSD), ignoring optical thickness effects. For HD172555, we have included the mass of SiO gas & large particle rubble in the estimate. For η Corvi, we quote the directly observed (by remote sensing) 0.1 – 100 μm mass in the abstract and use this quantity for determination of the minimum mass delivered.

## 4.1   Giant Impact Silica Production

In the Lisse *et al.* 2009 analysis of the HD 172555 system, we argued for a giant impact sourcing the system's silica debris disk (since imaged as predicted by deFalco *et al.* 2010 and Engler *et al.* 2018). We found a disk full of very fine (dn/da ~ a$^{-4.0\ +/-\ 0.1}$)   silica + olivinaceous dust, the expected result from a rocky body-rocky body collision happening at speeds great enough (> 10 km/s) to flash vaporize and reform siliceous dust. Giant impacts are thought to be common in the final stages of a planet's growth cycle (Kokubo & Ida 1998; Kokubo & Genda 2010; Kenyon & Bromley 2006). The energy required to make ~$10^{21}$ kg of fine silica dust + $10^{22}$ kg of SiO gas is readily available from the kinetic energy of collision of two Mercury sized bodies moving at the ~ 17.5 km/s Keplerian speeds at 6 AU in that system. Correspondingly in this system, there were almost no pyroxene silicates, but abundant olivine silicates (of both Mg-and Fe-rich varieties), and abundant silica alteration products. The preferential removal of ices, metal sulfides, and pyroxenes from the presumably originally primitive material occurred because these species are the least refractory present, and would not be stable in high temperature condensates formed from giant impact ejecta. SiO gas is easily formed via silicate impact vaporization. The net Fe:Si ratio was about the same as for primitive cometary material, suggesting Fe had not been preferentially removed from the rocky material during the alteration process. Creation & distribution of planetary-mass amounts of fine dust and boulders into circumstellar orbits was produced by the impact process ejecta. A very steep PSD is expected from a hypervelocity shock impact process at speeds great enough (> 10 km/s) to flash vaporize and reform siliceous dust, with dn/da ~ $a^{-4.1}$ (Takasawa *et al.* 2011).

A giant impact energy reservoir could potentially also be present for bodies at ~3 AU around the HD 145263 1.4 $M_{solar}$ F4V primary star, with ~21 km/s Keplerian velocities (~4.1 year circular orbital period). Young systems like the ~11 Myr HD 145263 system will have dynamically hot and unrelaxed planetesimal orbital distributions, meaning that hypervelocity impacts between planet-sized objects on orbits with very different inclinations are still possible. Relatively unrefractory materials like water ice, organics, and metal sulfides are absent, as expected for a thermally driven giant impact. **But the silicate abundance evidence is not as thermally predicted,** and the HD 145263 silica-rich MIR emissivity spectrum looks structurally very different from HD 172555's (Fig. 3). Instead of high temperature, refractory olivines, we find abundant low temperature pyroxenes, selective removal of Fe-olivines, and destruction of amorphous olivine in favor of amorphous pyroxene (Figs. 4 & 5). The PSD slope index we find, $a^{-3.75 +/- 0.1 (2\sigma)}$, is shallower than the $a^{-4.0 +/- 0.1 (2\sigma)}$ we found for the HD 172555 giant hypervelocity impact system, and closer to the dn/da ~ $a^{-3.5}$ to $a^{-3.7}$ expected for a debris disk containing dust from a collisional cascade with mass-independent or mass-dependent catastrophic disruption thresholds (Takasawa *et al.* 2011).

## 4.2    Giant Stellar Flares – Silicate Melting/Evaporation & Quick Refreezing

A silica producing mechanism involving a giant stellar flare or CME emitting tremendous amounts of optical light has the advantage that it could easily alter the circumstellar dust surrounding HD 145263 *in situ*. The flare/CME energy required can be calculated as follows. With a surface area of dust ~10 times that of HD 172555 (0.4 Jy @ 10 μm @ 142 pc vs 1.0 Jy @ 10 μm @ 29 pc) and a somewhat shallower PSD slope  (dn/da ~ $a^{-3.75}$ vs dn/da ~ $a^{-4.0}$ for HD 172555), the amount of mass in the HD 145263 disk is on the order of $10^{23}$ kg, 10 times greater than the HD 172555 disk. Using values of $\Delta H_{melting, basalt}$ = 500 kJ/kg, and  Heat Capacity$_{Basalt}$ ~ 1.2 kJ/kg/K, we find it takes 1.4 MJ/kg to melt rock at 1000K, and, based on the Silicon $\Delta H_{vaporization}$ = 383 kJ/mol * 35.7mol/kg = 13700 kJ/kg and an Si-O bond energy (452 kJ/mol) twice that of the Si-Si bond (222 kJ/mol), ~30 MJ/kg to destructively vaporize it (compare this to the 50MJ/kg estimate from Gail 2002 used in Lisse *et al.* 2009). Thus we need some $10^{29}$ to $10^{30}$ J to alter HD 145263's circumstellar dust, and this is assuming that all of the energy of a flare is directed into the flattish dust disk, which subtends only a few percent of the vertical solid angle of emission from the central star.

It would thus take some sort of extraordinarily huge and quick stellar mega-flare to rapidly deposit enough energy to heat the dust to boiling temperatures. Such a flare should be easily detectable from the Earth – and some are. According to Davenport 2016 and Van Doorsselaere *et al.* 2017's flare studies of Kepler/K2 stars, about 1.6% of such young F-stars flare detectably in the optical, but they can superflare with energies up to 3 x $10^{31}$ J. These super flares are very brief, on the order of 1–2 hours in duration, and occur every $10^4$ to $10^5$ days, so one of them could easily be missed by AAVSO-like variable star surveys observing on timescales of days to weeks.

In this scenario, HD 145263's circumstellar disk material was transformed on very short timescales of hours ($10^4 - 10^5$ s) sometime pre-2001 (the start of the ASAS monitoring of the system,, section 2.1), with melting and vaporization to significant depth accompanied by a huge impulse in radiation pressure. SiO gas was naturally formed via fast vaporization of silicate dust, and silicas by quick refreezing of the vapor. Removal of all pre-existing sub-blowout sized dust occurred, combined with a new particle size distribution created by equilibrium re-condensation and subsequent collisional grinding.

While this mechanism can explain the collisional-equilibrium like PSD we find for HD 145263's disk dust from the Spitzer IRS spectra (Sec 3), **a major discrepancy between the predictions of this mechanism for HD 145263's dust disk transformation and the observations are the presence of low melting point pyroxenes and the absence of crystalline Fe-olivine in HD 145263's disk dust (Figs. 4 & 5).** This combination of materials does not agree with the predicted thermal destruction of the least refractory ice, metal sulfide, and pyroxene species typically found in primitive PPD dust.

A variant on the giant stellar flare theme of note is a giant stellar flare induced by a giant impact onto the primary star. As has been suggested by the recent work of Brown (2016, 2018) in studying the sungrazing comet phenomenon in our solar system, a comet coming to perihelion in the Sun's photosphere moves with a velocity ~ 600 km/s with respect to the Sun, so that even a medium sized comet of $\sim10^{13}$ kg can deliver on the order of $4 \times 10^{24}$ J to the Sun upon impact. A medium sized asteroid or KBO massing $10^{18}$ kg or more could easily deliver $\sim10^{30}$ J to the HD 145263 primary's photosphere. What happens after that is pure conjecture; we do know from FU Ori stars that violent energy outbursts with short rise times seem to accompany massive accretion events, although settling times are usually much longer than 1–2 years and leave behind a tell-tale residual disk and its spectral signatures (Hartmann & Kenyon 1996; Green *et al.* 2006, 2016; Hubbard 2017). Singleton accretion events such as we are describing should happen most often in transition systems that have lost their primordial disks but have yet to clean out their planetesimal populations, and the low age of ~11 Myr for HD 145263 suggest that this could be plausible, although flares of this magnitude have yet to be observed in ~10 Myr old stars.

## 4.3   Space Weathering

Another variant on the stellar flare scenario of Section 4.2 that preserves collisional equilibrium particle shape, occurs at lower temperatures, requires much less energy, and acts selectively on Fe-silicates is the processing of circumstellar dust by stellar XUV and high energy particles to alter the surface layers of the circumstellar dust down to a few mid-IR skin depths, deeper than

~30 um. This can be done slowly, over millions of years, as in the exposed hard surfaces of bodies in our solar system, or more rapidly in the event of very large flaring events containing a similar equivalent total energy dose. If done quickly, the sputtering effects of stellar XUV and high energy particles can combine with thermal effects to efficiently transform the surface layers of a rocky object, mainly by removing oxygen from the tetrosilicate lattice and producing nanophase Fe from $Fe^{+2/+3}$ counterions.

How does this scenario match up to what was observed in HD145623? The observed mineralogical mix does not match up with any known spontaneous terrestrial chemical alteration of primitive cometary rocky material, nor the species found in aqueously altered asteroids. Aqueous alteration of silicates would ***oxidize*** the $Fe^{+2}$ in the silicates, producing olivine and silica from pyroxene, along with phyllosilicates, carbonates, and Fe-oxides. We see none of the latter 3 materials in the Spitzer IRS mineralogy. ***This rules out the possibility of alteration to silica within compact bodies that this dust may have been sourced from.*** Flash vaporization and recondensation of rocky material would create a mixture of high temperature stable and kinetically favored products, i.e. olivines + silicas, as we saw in the HD172555 disk system mineralogy, where pyroxenes had been almost totally destroyed (Figure 4). ***Since we do see abundant Fe-pyroxenes in the HD145263 disk dust (Table 3), we can rule out a hypervelocity impact formation process for the dust.*** Instead, in HD 145263 the silicates have been ***reduced*** from a predominantly olivine composition to a predominantly pyroxene + silica composition, although Mg-olivine (forsterite) apparently remains unaltered.

This observed pattern of mineralogical change ***is*** consistent with low level heating of primitive, mostly amorphous material. Temperatures would have to have been high enough to completely vaporize water ice and metal sulfides, while creating an appreciable vapor pressure of recondensing SiO. At the same time, temperatures would have to have stayed low enough to not alter the Mg olivines, implying T < 1300 K (Nuth & Johnson 2006). Depletion of O, C, Mg, and Fe could have occurred via this mechanism, as O, Mg, and Fe become volatile above 1000 C in tetrosilicates and the released O can attack the refractory C component, producing volatile CO and removing some of the C from the system. (One has to be careful about estimating the total system Fe, Mg, and O abundances because, while our NIR spectral allow us to rule out the presence of any appreciable circumstellar Fe/Mg atoms or low ionization state ions, there could be significant amounts of these atoms in hard to detect recondensed Fe- and Mg-oxides.) Sputtering alteration of the dust by energetic particles in HD 145263's stellar wind could also play a role, with Fe-olivines again most easily reductively transformed in their optical properties (Quadery *et al.* 2015).

Moderate heating + energetic particle sputtering of dust in vacuum goes by another name in planetary astronomy – "space weathering", and is a keen area of study for interpreting the different surface reflectance spectra of the ~$10^6$ asteroids in our solar system (Domingue *et al.* 2014; Fazio *et al.* 2018; Kohout *et al.* 2014, 2016; Pieters & Noble 2016; Reddy *et al.* 2012; Quadery *et al.* 2015; Schelling *et al.* 2015). These studies generally show that of the common ferromagnesian tetrosilicates, it is easiest to weather Fe-bearing olivines into amorphous pyroxenes, silica, and Fe-metal via removal of Fe and O from the surface layers of the olivine. A critical point to add is that the amount of energy required to space weather alter the optical properties of primitive era HD 145263 circumstellar dust to appear like the present era material to a few skin depths (~30 um) can be much lower than that required to totally vaporize and reform it, especially for the biggest dust grains. Assuming the net reaction is something like $8H + 3 FeSiO_4 \rightarrow Fe_2SiO_6 + 4 Fe + SiO_2 + 4 H_2O$, the total change in Gibbs free energy is (-267.2 - 204.6 - 4*54.64 + 3*329.6)*4.184 = +1250 kJ/(3 moles of transformed fayalite), or ~2.1 MJ/kg of transformed fayalite (assuming 5 mol of fayalite/kg of rock). This is 15 to 25 times less energy/kg than the giant flare dust vaporization scenario of Section 4.2 requires, since the "low-level" heating to 1100–1300 K and stellar wind sputtering does not require bulk destructive vaporization of all of the circumstellar dust. If we further constrain the transformation to depths of ~30 $\mu m$, only about 5% of the dn/da~ $a^{-3.75}$ (Section 3) dust mass needs to be space weather altered, and the total flare energy required is reduced by a factor of 300 to 500 versus the $10^{29} – 10^{30}$ J we estimated in Section 4.1 would be required to vaporize the dust in place. I.e., we estimate a total energy on the order of $10^{26} – 10^{27}$ J ***is*** sufficient to transform HD 145263's dust from its primordial state to its present state. This is comfortably within the range of the larger stellar XUV flares seen in the Milky Way (Byrne 2012).

There is no listed x-ray detection or information about HD 145263's x-ray luminosity in the archival literature; it was not detected in the ROSAT All-Sky Survey (RASS). We gain some assurance that an ~12 Myr old F4V star like HD 145263 should be XUV and stellar wind active using the Chandra values of $L_x = 2 \times 10^{21}$ W for ~12 Myr old F2V star HD 113766A and $L_x = 2 \times 10^{22}$ W for ~12 Myr old F6V star HD 113766B at d ~ 120 pc (Lisse *et al.* 2017c), and by noting that stellar x-ray emission and flaring activity increases markedly as one goes from early to late F-type stars. (A rotational period of 1.7 days for the HD 145263 primary (Section 2) implies a Rossby number of ~0.07 (Wright *et al.* 2011), and a star rotating rapidly versus its convective timescale in the $L_x = 0.0006 \ L_{bol}$ saturation regime. Using $L_x = 0.0006 \ L_{bol}$ , we would estimate an even higher $L_x = 3 \times 10^{23}$ W.) Assuming that it is active, we can make an educated guess at its level of x-ray activity (or that of any nearby optically undetected M-type companions) using the RASS limiting sensitivity of ~ $3 \times 10^{22}$ W at d = 142 pc. Since brief giant flares of up to 1000x the quiescent emission level are not uncommon for young stars (Ribas 2005, Guinan & Engle 2007, Osten & Wolk 2015), we can assume a most optimistic HD 145263 case with a giant XUV flare energy flux of ~ $3 \times 10^{25}$ W, some 3 orders of magnitude lower than the maximal optical flare flux presented in Section 4.1. It would take $3 \times 10^4$ to $3 \times 10^6$ s (or 10 hrs to 1 month, with the latter longer estimate assuming that ~5% of the radiated XUV energy is directed into the disk and not into $4\pi$ of free space), to accumulate the ~ $10^{27}$ J of XUV energy observed in the largest stellar magnetospheric flares/CMEs ever recorded (Byrne 2012; by contrast the largest flare/CME total energy ever observed from the modern, 4.56 Gyr old Sun is ~ $10^{26}$ J in the Carrington event of 1856 and the great storm of 2012). A series of closely spaced (in time) flares could also produce the observed result, as long as the time required to achieve the total flare energy dosage is short compared to any patina removal timescales, which are on the order of decades (see Section 4.4).

Thus an XUV flare irradiation mechanism could produce the olivine -> pyroxene + silica mineralogical transformation seen in HD 145263's disk, while providing enough energy to transform an optically thick surface patina on the disk's dust. From all the evidence, then, the best explanation for what has occurred in HD 145263 is that we have encountered a severe case of space weathering alteration of a young debris disk. If so, then the findings we have from studies of space weathering in our solar system can be applied to HD 145263's disk material – e.g., we can expect the dust to be optically reddened vs the usual grayish neutral colors of silicates, and for there to be abundant nanophase Fe present.

## 4.4  Summary & Need for XU/VIS/IR Monitoring of Silica Debris Disk Systems

In summary, the only physical mechanism we currently know of that could produce the observed HD 145263 silica-rich dust mineralogy from an initially primitive dust reservoir would be space weathering caused by an epoch of unusually intense stellar wind or a giant stellar/XUV outburst from the central star. Unfortunately for this hypothesis, however, HD 145263 is not a detected source in the 2nd ROSAT All-Sky Survey catalogue (Boller *et al.* 2016), which adds little credence to the "active young star transforming nearby circumstellar dust" by frequent flaring over time scenario. (On the other hand, we do not expect the average young mid-F star at 142 pc from the Earth and ROSAT, with log $L_x = 28.9 - 29.5$ to be detectable in the RASS (Panzera *et al.* 1999, Favata & Micela 2003, Schmitt *et al.* 2004, Suchkov *et al.* 2003) and the star appears to be rotating rapidly enough with a 1.7 day period (Fig 1b-c) to be XUV active (Section 4.2).) Nor do the AAVSO or superWASP photometric time series for the star in the 2003–2009 timeframe, which show no excursions larger than ~0.2 mag in apparent brightness from the mean (Fig 1b, 1c), help advise us. [Again on the other hand, giant stellar XUV flares often do not have optical counterparts (E. Shkolnik 2018, E. Guinan 2019 priv. commun.), and can be of very short duration and so missed by weekly monitoring photometry (Balona 2012)].

Because we only rarely see silica-dominated debris disks, we are driven towards assuming extreme space weathering due to some sort of unusually strong stellar wind or a stochastic, quick superflare in the system, but have no direct proof of this. A superflare disk transformation mechanism should be testable by high cadence optical and X-ray lightcurve monitoring of the system. One of the near-future methods of doing this would be to examine the 2-minute cadence TESS lightcurve data for this star for flares. At apparent visual magnitude V ~ 9, HD 145263 is comfortably above the TESS limiting magnitude of V~12, and any photometric variations due to flaring should be detectable. Another indirect test of the space weathering hypothesis is to measure the mineralogy of the HD 145263 disk over time, to see if the surface patina of reduced pyroxenes + silica wears off over a relatively short (in geologic & orbital terms) amount of time, as opposed to the Myrs estimated to replace flash refrozen dust from a massive hypervelocity impact (Jackson & Wyatt 2012).

There is a useful additional calculation we can make regarding timescales here. Any newly space-weathered dust should start being depleted as soon as it is created by the forces acting most readily on the small circumstellar particles, radiation pressure and P-R drag. In a collisionally supported dust population, however, quickly removed small particles can always be replaced by grinding down of longer-lived bigger particles, as long as they are present and contain significant amounts of space-weathered material to be ground. As large particles for which an ~30-μm skin of space weathered material is only a small fraction of the mass begin to grind down, they will replenish the original comet-like dust composition. The breakpoint size is about 0.3 mm diameter, the size at which a 30-μm altered skin would account for more than half the mass of a particle. This suggests that the evinced disk composition should trend back towards the original cometary composition on a timescale that is roughly the collision time for 0.3 mm particles in the HD 145263 disk. The collision time is roughly $1/(n\sigma v_{rel})$, where $n$ is the number density, $\sigma$ is the collisional cross-section, and $v_{rel}$ is the relative velocity. If we assume a disk width/radius ratio of about 0.1, for a disk width of 0.3 AU and all of the $10^{22}$ kg of sensible dust in 0.3 mm dust grains with a density of 3.0 g/cm$^3$ (Tables 3, 5) then we have $n \sim 3 \times 10^7$ m$^{-3}$. If we also assume that the relative collision velocity is about 10% of the orbital velocity, and the collisional cross section is the dust hard sphere cross section for 0.3 mm diameter spheres, then that gives us a lower-limit to the collision clearing timescale of about 10 years, or about 2.5 dust orbits around the primary. Since we do not see any appreciable change in the Subaru/COMICS 8–13 μm spectral character of the dust from 2003 through 2014, we can presume the dust has survived at least the expected minimum amount of collisional grinding time. But it is very possible that in the 5 years since the last *COMICS* observation that the evinced dust surface composition could have started going back towards its primitive composition, and that future observations in the 2020's, e.g. with JWST, or with ground-based mid-IR spectrometers currently available, should show this. A collisional support timescale on the order of 10 years also explains why space weathering effects are so ephemeral vs geological and solar system evolutionary timescales – severely altered space weathered surface material is not well supported by the large particle reservoir for very long.

Further monitoring of the HD 145263 system in the MIR and X-ray is thus clearly warranted, to see if it does indeed occasionally superflare, and if its dust has only been altered in a 10's of microns surface patina. The same can be said for the HD 172555 system. In comparison to our new HD 145263 result, the high temperature mineralogy and extremely fine-grain dominated dust grain dn/da ~ $a^{-4.1}$ PSD we see in the HD 172555 silica rich disk today **could not** have been formed via relatively gentle, shape-preserving space weathering. The transformation at T > 1300 K (to destroy all pyroxenes, and flash freeze new silica) would have to have been quick and severe, requiring incredibly high power levels in any superflare - potentially like one of the ~$10^{31}$ J optical superflares described in Section 4.2.

The planet-planet hypervelocity collision scenario we promulgated for producing the abundant silica found in the HD 172555 disk (Lisse *et al.* 2009) was determined using a process of elimination of all known mechanisms for its creation. It still seems highly likely as a probable cause of the rare silica dominated debris disks, as they are found in young systems in their planet building phase and planets are thought to grow to their final sizes via giant collisions (Kokubo & Ida 1998; Kokubo & Genda 2010; Kenyon & Bromley 2006, Jackson & Wyatt 2012). But the existence of such high intensity, short duration stellar flares was not known until the Kepler/K2 stellar survey work of 2012 – 2017 (Davenport 2016 and Van Doorsselaere *et al.* 2017), and the new TESS work in 2019 (Gunther *et al.* 2019). Even though young, ~20 Myr old A7V stars like HD 172555 are seen to superflare orders of magnitude less often than FGK stars, we need to add this new possibility of an optical superflare totalling some $10^{30}$ - $10^{31}$ J energy transforming a pre-existing primitive dense dust disk to the list of possible causes.

This new possibility should be tested by new high cadence optical monitoring. If optical superflares are found from the HD 145263 primary, it will show that there are possibly 3 ways to make a silica-rich debris disk (via optical superflares, XUV space weathering, or giant impacts). This will have important implications for how the silica around 25% of T Tauri stars, and the silica found in solar system meteorites was made (Sargent *et al.* 2006). We won't know for sure, though, unless a future superflare is seen coupled with another MIR spectral silica transformation event.

## 6. Conclusions

In this paper we have presented spectra and photometry of the HD 145263 debris disk system over the 2003 – 2014 timeframe. We find an F4V host star surrounded by a stable, massive $10^{22}$

– $10^{23}$ kg dust disk equivalent to the material found in a Moon- to Mars-sized body. No gas was found in the circumstellar disk, and the primary star was found to be rotating at a rapid rate, with an ~1.75 day period. The silica + pyroxene rich mineralogy of the dust disk is very unusual compared to the ferromagnesian silicates typically found in primordial and debris disks.

Using our previous study of the young HD 172555 silica rich debris disk (Lisse *et al.* 2009) as a guide, we raise a number of possible silica formation scenarios for HD 145263. But unlike that study, we have additional information: the time series of mid-infrared spectra straddling the appearance of circumstellar silica dust, and a medium resolution near-infrared NASA IRTF/SpeX spectrum that allows us to rule out any stellar accretion, strong wind outflow, or unusual system mineralogical abundances. The same spectra do not show any SiO emission features in the 4–5 μm $1^{st}$ overtone region, supporting our new supposition that the 8 μm "SiO gas feature" is explainable instead as a hydrated silica solid state feature. Another important clue is produced by our spectral modeling: if we take a typical protoplanetary dust disk mineralogical mix, and delete the Fe-rich olivines, water ice, and metal sulfides, and replace them with amorphous and Fe-rich crystalline pyroxene and silica, we can reproduce the Spitzer spectra. Putting this all together, we find that the only mechanism we can propose that is energetically reasonable for reductively altering an orbiting dust population in place as seen is space weathering triggered by giant stellar flare transformation of the Fe-silicate components of the original cometary dust population.

Conversely, re-examining the Lisse *et al.* 2009 HD 172555 observations, we find a very different circumstellar dust population in that system, one that is heavily dominated by fine silica dust. The minor constituents contain abundant silicate and metal-sulfide crystalline materials but little amorphous components. The HD 172555 population looks like a thermally annealed dust disk to which a huge amount of new glassy silica dust has been added. So a giant hypervelocity impact in this system, as we proposed in 2009, still seems to be a viable option. The work discussed here has added the possibility of a giant optical stellar flare melting and vaporizing a pre-existing dust disk.

In sum, there may be up to 3 ways to make silica rich debris disks, an idea consistent with the 3 qualitatively different kinds of silica disk MIR spectra seen in Figure 2a (HD 145263 + HD 166191; HD 172555; HD 23514 + HD 15407A), and that spectrally monitoring the known silica disks on yearly and decadal timescales would be useful to help discern the mechanism(s) of their production.

For HD 145263, with its space weathered dust disk, we predict that there should be copious amounts of optically red, nanophase Fe-rich dust in the disk orbiting around the primary. We also predict that the space weathering signature should disappear over time if the dust was truly altered only to a shallow 10's of microns depth, as the altered small particles are swept out of the system and grinding of larger particles to replace them exposes fresh, un-space weathered dust. This should happen over collisional and radiative blowout timescales. It would then take another stochastic super flare from the star to re-weather material and re-instate the silica + amorphous pyroxene appearance. It is thus important to obtain future high cadence optical and X-ray lightcurve monitoring of the system.

Finally, there is still the mystery of how $10^{22} - 10^{23}$ kg of dust became emplaced at ~3 AU from the HD 145263's F4V primary to begin with. The best explanation we can find is that the $10^{22} - 10^{23}$ kg of dust is some of the last remnants of the HD 145263 system's original primordial disk, and that we are seeing a transition disk's processing (Section 2), but this needs to be proven by future work. This line of inquiry has important implications for the creation of silica found in abundance orbiting around ~25% of T Tauri stars and in some primitive solar system meteoritic materials (Sargent *et al.* 2006), and needs to be pursued in order to better understand early solar system development.

## 7.    Acknowledgements

This paper was based on observations taken with the NASA Spitzer Space Telescope, operated by JPL/Caltech, on observations taken by the Subaru Telescope, funded and supported by the National Astronomical Observatory of Japan, and on observations taken with the IRTF/SpeX 0.8–5.5 Micron Medium-Resolution Spectrograph and Imager, funded by the National Science Foundation and NASA and operated by the NASA Infrared Telescope Facility. The authors would like to acknowledge help in obtaining the data from H. Fujiwara, M. Honda, and J.

Rayner, and highly useful comments and suggestions received from colleagues S. Engle, E. Guinan, and D. Watson used in forming the analysis & discussion presented in this paper. The authors would also like to acknowledge the very important role that referee M. Perrin's constructive reviews played in making this paper a robust reality. C. Lisse gratefully acknowledges support for performing the modeling described herein from JPL Spitzer contract 1274485, NASA NExSS grant to ASU NNX15AD53G, and NSF Grant AST-0908815. M. Lucas was supported by a scholarship from the NASA Iowa Space Grant Consortium.

## 9. Figures

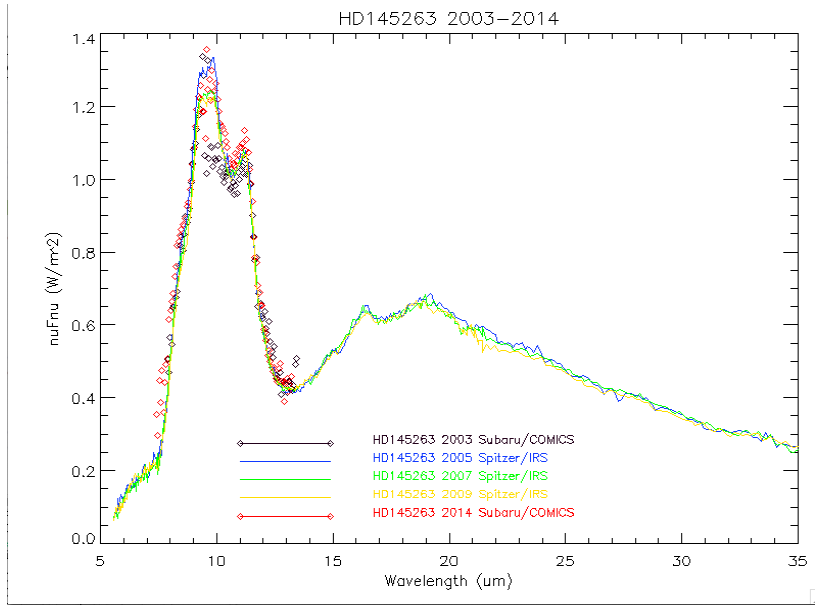

**Figure 1a** - Comparison of the revised 2003 Subaru/COMICS HD 145263 observation in 2003 (originally published in Honda *et al.* 2004 and corrected in this work; **black diamonds**) to the Spitzer/IRS spectra from the 2005 through 2009 epochs (blue/green/yellow). For clarity, error bars are not plotted. Also shown is the Subaru/COMICS HD 145263 observation of Fujiwara *et al.* (2013; red diamonds). Error bars have been suppressed to allow direct comparison of the agreement between the different spectra. The errors in an individual visit can be estimated from the scatter in the data, and the relative errors between repeated visits from each telescope average 10%. The absolute calibration difference between the Subaru and Spitzer datasets averages 30% due to differences in the effective beamsize of each slit, with the Subaru measured Jy being 1.3 times higher than the reported Spitzer spectral fluxes. After correcting the Subaru data to match the Spitzer calibration as shown. the total $8 - 13$ µm emitted radiance for the system is found to vary by less than 10%.

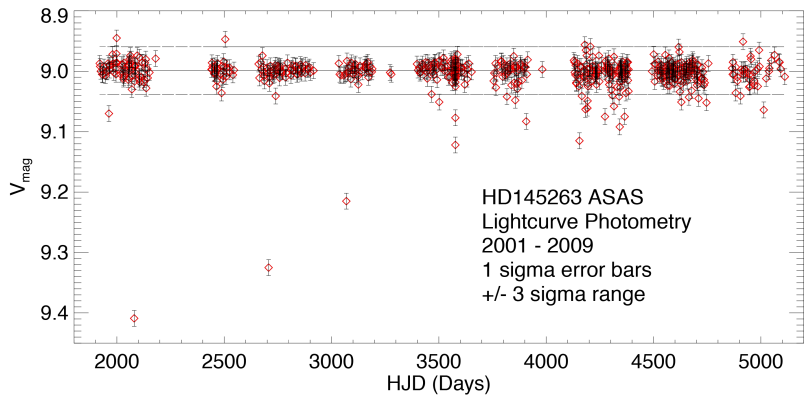

**Figure 1b** - Long term V-band photometry for HD 145263 from the ASAS website ([http://www.astrouw.edu.pl/asas/?page=aasc](http://www.astrouw.edu.pl/asas/?page=aasc)) spanning 01 Jan 2001 (HJD = 2451910.5) to 10 Oct 2009 (HJD = 2455105.5). Subaru/COMICS observed the system on JD 2452835.5 (07/15/2003) and the first Spitzer/IRS observation occurred on JD 2453444.5 (03/15/2005). There are 3 single point measurements in this interval differing by somewhat more than $3\sigma$ from the V=8.99 temporal median that would be indicative of long term stellar flaring; for 622 measurements, we expect 1.7 for normally distributed statistics. There are multiple epochs of single point excursions much fainter than the V=8.99 temporal median by up to 0.40 mags, but they are not reproduced by neighboring samples and it is not clear whether they represent real changes in the system's visual output or simply noisy data dropouts.

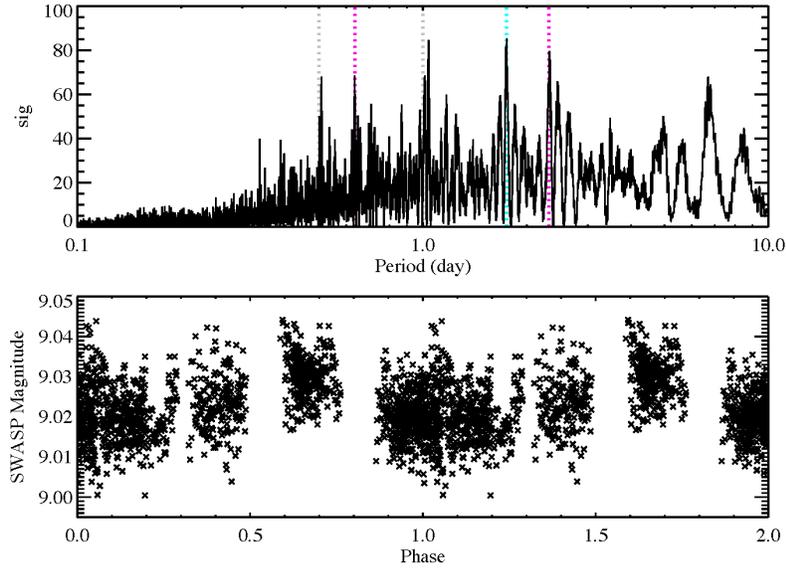

**Figure 1c** - SigSpec periodogram of the filtered superSWASP time series obtained between 05 Jul 2006 and 08 Aug 2008. (Top) The power spectrum. The strongest signal has a period of 1.747 ± 0.001 days, with aliases peaking near 0.64 and 2.31 days. Artifacts associated with the observational cadence can be seen near 0.5, 1.0, and 6.67 day. (Bottom) The HD145263 superSWASP phase curve folded with the 1.747-day period.

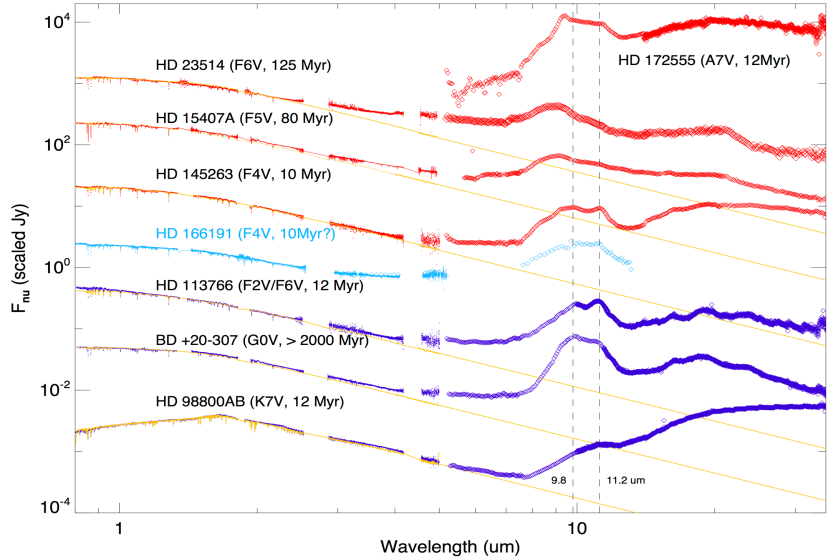

**Figure 2a –** Comparison of the 0.8 − 35 μm SpeX + 2009 IRS SED for HD 145263 to that of 7 other warm dust systems (3 silica dominated, in **red**; 3 silicate dominated, in **dark blue;** and 1 transition disk, in **light blue**). The MIR excesses of the silica systems have excess flux signatures extending to shorter wavelengths (7-9 μm) than the silicate disks (9-11 μm), as the silica SiO fundamental stretches are at 8-9 um. The HD 172555 IRS spectrum at upper right shows the strongest and most pronounced silica dust emission features. HD 23514 and HD 15407A show smooth excess emission features mainly dominated by silica dust; HD 113766 and BD+20-307 show the multiple peaked 8-13 μm structure typical of mixed olivine + pyroxene silicate dusts. HD 145263 is a hybrid mix between the two, evincing multiple 8-13 peaks as well as a pronounced 7-8 μm shoulder. Transition disk HD 166191, only observed using ground based observations (Kennedy *et al.* 2014), looks very similar to HD 145263. HD 98800AB is silicate rich, but the MIR SED is dominated by a huge amount of T~150K long wavelength large dust particle blackbody emission seen as a minority contributor in HD172555 and HD 145263. All 7 of the IRS spectra show significant excess out to the 35 μm Spitzer/IRS cutoff, consistent with their being abundant dust particles with radii > 30 μm present.

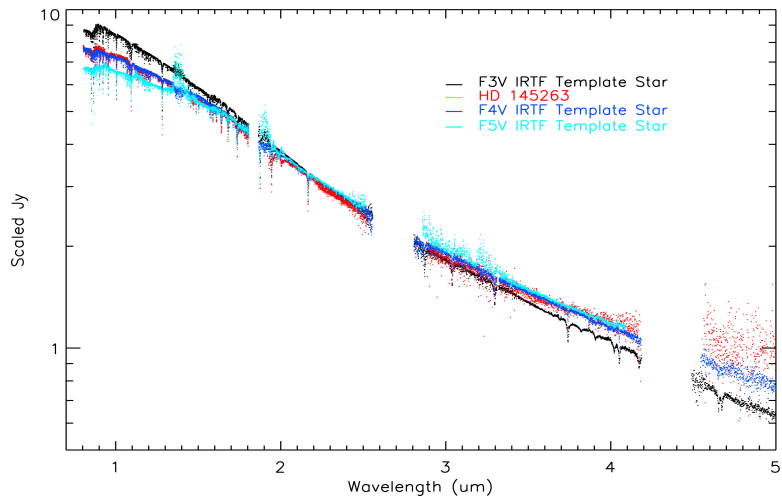

**Figure 2b –** HD 145263's 1-4 μm spectral structure best matches that of an IRTF spectral library archival template star F4V star, quite different from previous descriptions of the star's spectral type as F0V (Sylvester & Mannings 2000, Honda *et al.* 2004).

**Figure 2c** – Detailed view of the HD145263 IRTF/SpeX spectrum in the 1.9 – 2.4 μm wavelength range, showing the lack of CO, HeI, FeII, HBrγ, H₂ S(0), and SI/SrI/SiI emission lines as seen in the spectra of other NIRDS targets, like NovaOph, 51Oph, and HD166191. There are no spectral earmarks due to circumstellar gas in this system, nor is there any evidence of high temperature processes like photospheric accretion, wind outflow, or dust sublimation. The Fe/Si and Mg/Si EW ratios are also similar to those for other early solar abundance F-stars in our sample, ruling out unusual, non-solar abundance patterns producing the observed dust mineralogy.

**Figure 2d & 2e - SpeX 0.96 – 1.07 and 1.48 – 1.75 μm absorption line spectra of HD145263 vs NIRDS spectra and "stellar template"** (i.e., IRTF/SpeX on-line stellar library archival spectra) **spectra** of matching stellar type. Major clean (i.e., unconfused) atomic absorption lines are marked by vertical dashed lines in both plots. **(d)** Shows that HD 145263 primary's absorption features are much more consistent with the deeper and more structured lines of the mid-F stars (e.g., at 0.988, 1.042, 1.068, and 1.083 um) in our NIRDS sample than for the previous literature F0V assignment for the star. **In (e), we show the HD145263 – F4V correspondence for the 1.49 – 1.75 μm range**. HD145263 has been matched to an F4V template star's spectrum from the NASA/IRTF stellar library. **Spectral comparisons for 2 other silica dominated debris disk systems are also shown - HD15407A** has been matched to an F5V star's spectrum (green curve), and **HD23514** has been matched to an F6V spectrum (blue curve) and a G0V star's spectrum (aqua curve). **In d & e**, there is no obvious difference between the spectra of silica disk hosting stars and the non-hosting template stars in this highly diagnostic spectral range, arguing against any unusual primordial system abundances causing the observed silica rich dust.

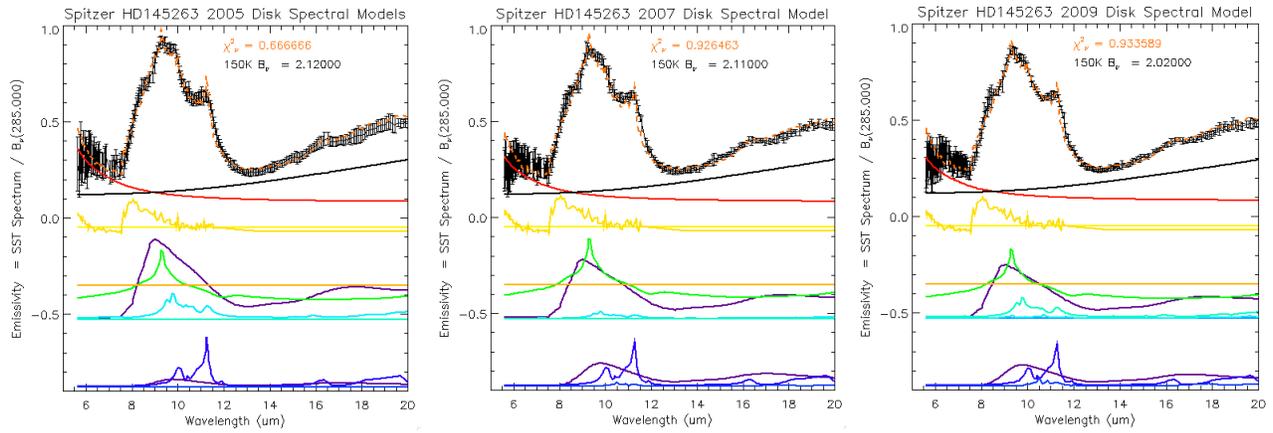

**Figure 3 – Mineralogical model decompositions of the Spitzer spectra. (a-c)** Shown are 3 panels corresponding to the three 2005 - 2009 epochs of Spitzer spectral observation of HD 145263. At top of each panel in black are the temperature-independent "emissivity spectra" (i.e., observed flux vs wavelength divided by a simple best-fit blackbody). The dashed orange line is our best-fit laboratory analogue emissivity model. Error bars are 2-sigma. Informed by our previous model fits to cometary, asteroidal, and debris disk spectra (Lisse *et al.* 2006, 2007a,b, 2008, 2009, 2012, Reach *et al.* 2010, Currie *et al.* 2011, Sitko *et al.* 2011) we tried linear combinations of 80+ different materials to see how well they would match the observed emissivity. The results quickly boiled down to a simple, refractory mixture of amorphous pyroxene (middle dark purple), amorphous carbon (red), amorphous silica (bright green), SiO gas/hydrated silica (yellow), crystalline olivine (forsterite, dark blue) and amorphous olivine (lower dark purple) and crystalline pyroxene (ferrosilite, light blue). The amplitude of each colored curve denotes its relative abundance in the fit; for example, the dark purple amorphous pyroxene abundance is ~2x more abundant in 2005 than it is in 2009. The same basic recipe fit all 3 epochs of Spitzer spectra (orange dashed curve), with relatively more amorphous pyroxene and carbon in 2005 and more amorphous olivine in 2009. Of greatest importance is the mineralogy: the 2005 - 2009 dust populations contain no crystalline Fe-olivine fayalite, Fe sulfides, crystalline pyroxene orthoenstatite or diopside or water ice, and very little amorphous olivine, while large reservoirs of amorphous pyroxene and silica are present, along with some ferrosilite, very different from any primitive dust mineralogy we have ever studied. **(d)** Reproduction of our best-fit spectral decomposition model from our 2009 HD 172555 study (Lisse *et al.* 2009, Johnson *et al.* 2012), with the color coding of the observations and individual components the same as in panels a) through c). This debris disk is overwhelmingly dominated by very fine silica dust, without any obvious amorphous pyroxene, and the minor silicate components are both Mg- and Fe-rich. Ferromagnesian metal sulfides are also present. The chemistry of this dust is very different than the HD 145263 disk dust.

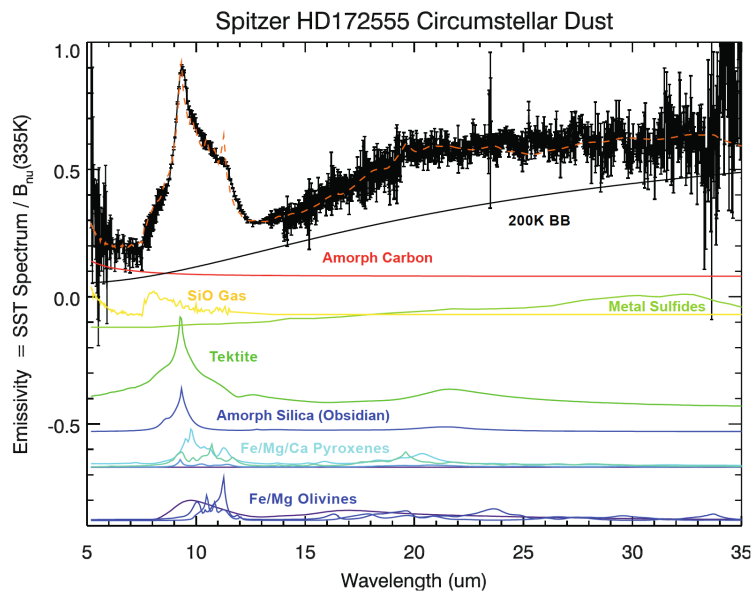

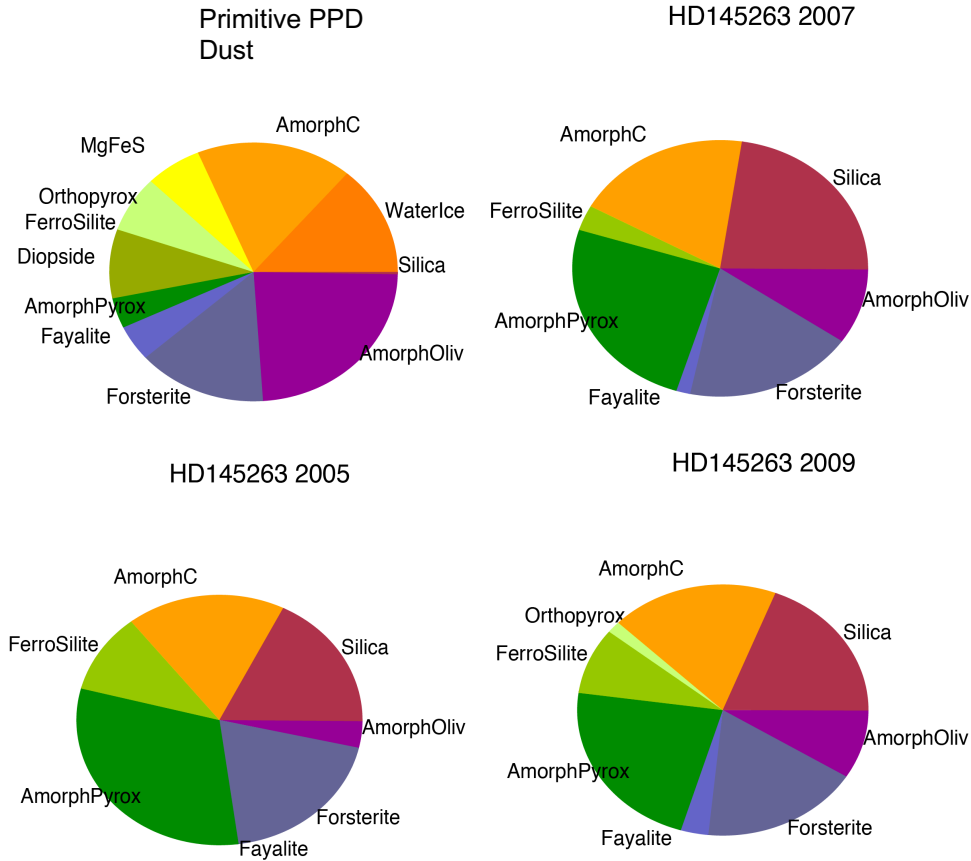

**Figure 4** – Pie charts showing the relative surface areas of the minerals detected in the circumstellar dust of HD145263. **(a)** Relative surface areas of the dust mineral components in a PPD system, as estimated using a comet-like mineralogy with no silica + abundant water ice, FeMg-sulfides and an amorphous/crystalline mix of FeMg olivines and pyroxenes (Lisse *et al.* 2006, 2007). **(b-d)** The markedly different mineralogical mixes of 2005-2009: Fe-olivine, FeMgS, and water ice are absent, while silica and amorphous pyroxene are superabundant. Crystalline Mg-olivine and amorphous carbon are found at levels expected for primitive cometary dust. The dust mineralogical mix is relatively stable from 2005 to 2009.

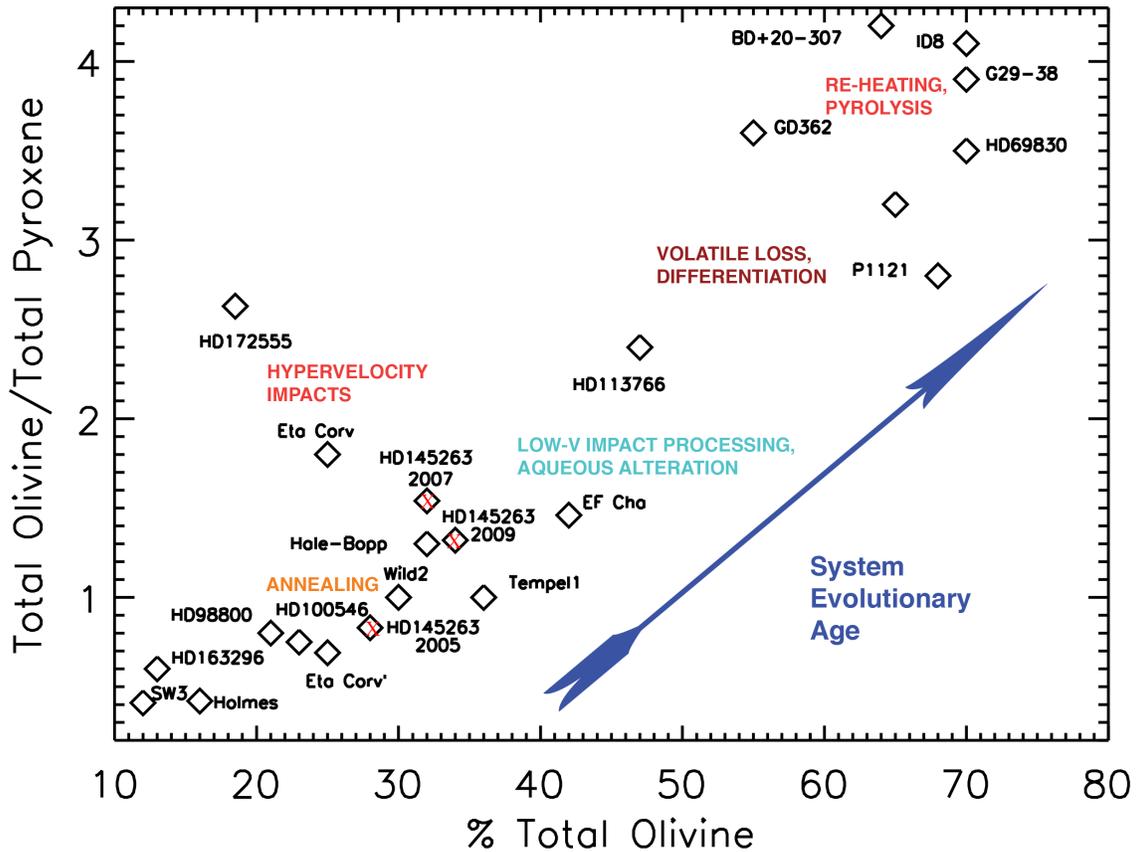

**Figure 5 — Silicate mineralogy for** the 3 epochs of HD 145263 *Spitzer* mid-IR observations described in the text versus that found in 5 comet systems (SW3B/C, Sitko *et al.* 2011; Hale-Bopp, Lisse *et al.* 2007a; Holmes, Reach *et al.* 2010; Tempel 1, Lisse *et al.* 2006; and Wild 2, Zolensky *et al.* 2007, 2008), the primitive YSO disk systems HD100546 (Lisse *et al.* 2007a) and HD163296; the primitive dust found around HD98800B; the primitive rock + ice dust found in the η Corvi LHB system (Lisse *et al.* 2012); the crystalline, S-type asteroid material in the ~10 Myr old F5V terrestrial planet forming system HD113766 (Lisse *et al.* 2008); the hypervelocity impact debris created in the 12 Myr HD172555 system; the mature asteroidal debris belt system HD69830 (dominated by P/D outer asteroid dust; Lisse *et al.* 2007b) and the similarly rocky material found in the hyper-dusty, solar aged F-star system BD+20-307 (Weinberger *et al.* 2011); and the ancient debris disk of white dwarfs G29-38 (Reach *et al.* 2008) and GD362 (Jura *et al.* 2007). The general trend observed is that the relative pyroxene content is high for the most primitive material (i.e., YSOs), and low for the most processed (i.e., white dwarfs). Young systems with material altered by high velocity impact processing, like HD172555 and η Corvi, lie above and to the left of the trend line. Note, that while the HD145263 points all lie on the trendline with olivine:pyroxene abundances consistent with primitive cometary material, the dust in these epochs also contains large amounts of silica which is not detected in comets. Some unusual process not connected with the normal thermal annealing of amorphous dust, oxidative aqueous alteration, and low velocity impact grinding seen in comets and debris disks must be at play in HD 145263.